%% file: main.tex
\documentclass[aps,pra,twocolumn,floatfix,superscriptaddress,nofootinbib,showkeys,10pt]{revtex4-2}
\usepackage{mathbbol}
\usepackage{amsmath,amssymb}
\usepackage{tikz}
\usepackage{centernot}
\usepackage[utf8]{inputenc}
\usepackage[english]{babel}
\usepackage{amsfonts}
\usepackage{amsthm}
\usepackage{mathrsfs}
\usepackage{mathtools}
\usepackage[nice]{nicefrac}
\usepackage{amsmath}
\usepackage{xfrac}
\usepackage{tabularx}
\usepackage{physics,subfigure}
\usepackage[colorlinks=true,hyperfootnotes=true,breaklinks=true,citecolor=black,urlcolor=black,linkcolor=black]{hyperref}

\usepackage{url}
\usepackage{enumitem}

\begin{document}

\title{Metrological Benchmarking of Random Quantum Circuits}

\author{Simone Cavazzoni}
\email{simone.cavazzoni@kaist.ac.kr}
\affiliation{Department of Physics, Korea Advanced Institute of Science and Technology, Daejeon 34141, Korea}

\author{Changhun Oh}
\email{changhun0218@gmail.com}
\affiliation{Department of Physics, Korea Advanced Institute of Science and Technology, Daejeon 34141, Korea}

\date{\today}

\begin{abstract}

Random circuit sampling is a leading approach to demonstrating quantum computational advantage, but benchmarking noisy implementations through linear cross-entropy requires costly calculations of ideal output probabilities. We propose a metrological benchmark based on the response to controlled perturbations, characterized by quantum Fisher information (QFI). A relation between QFI and out-of-time-order correlators enables protocols with local or global control. For Haar-random circuits, the average QFI approaches its maximal value in the local protocol and grows linearly with the number of qubits under collective control. In contrast, Clifford circuits yield zero QFI despite extensive operator spreading, showing that the response probes dynamical properties beyond spreading alone. A butterfly protocol further uses system size to enhance sensitivity, yielding a mean inverse sensitivity proportional to the number of qubits with single-qubit readout. Its fluctuations distinguish the Haar and Clifford ensembles despite their identical mean responses. For noisy implementations, we derive an exact relation between the noisy and ideal QFI within a global white-noise model, providing a quantitative reference for the degradation of the benchmark. These protocols enable metrological benchmarking of noisy random quantum circuits without computing ideal output probabilities.

\end{abstract}

\maketitle

Random circuit sampling (RCS) is a leading candidate for demonstrating quantum computational advantage~\cite{hangleiter2023computational}. Experiments aim to generate samples from quantum circuits whose output distributions are difficult to reproduce using classical computers~\cite{arute2019quantum,wu2021strong,zhu2022quantum,morvan2024phase,gao2025establishing}. However, noise can make these distributions easier to sample from classically, undermining the advantage sought in these experiments~\cite{noh2020efficient,aharonov2023polynomial,schuster2025polynomial}. Reliable benchmarks are therefore needed to assess the performance of noisy implementations~\cite{liu2021benchmarking}.

Linear cross-entropy benchmarking (XEB), widely used in RCS experiments~\cite{arute2019quantum,wu2021strong,morvan2024phase}, is evaluated using the ideal probabilities of experimentally sampled bit strings. Theoretical studies have established conditions under which XEB tracks circuit fidelity~\cite{dalzell2024random,ware2023sharp,gao2024limitations} and examined the classical complexity of achieving a given XEB score~\cite{aaronson2020classical,barak2021spoofing,gao2024limitations}. While XEB does not require reconstructing the full output distribution, computing the required ideal probabilities becomes expensive for large random circuits~\cite{boixo2018characterizing,hangleiter2023computational}. This computational cost limits the benchmark's scalability and motivates methods that do not rely on classical calculations of ideal output probabilities. A complementary experimental direction uses circuit dynamics to probe properties of random circuits. Recent experiments have used time-reversal protocols to measure higher-order out-of-time-order correlators (OTOCs), revealing many-body interference in random quantum circuits~\cite{google2025observation}. These experiments motivate using dynamical responses to probe circuit properties beyond the spreading of initially local operators.

In this work, we propose an experimentally accessible metrological benchmark for random quantum circuits. We characterize a circuit through the response of a quantum state to a controlled perturbation, rather than through its output probabilities. We derive a relation between quantum Fisher information (QFI)~\cite{paris2009quantum} and OTOCs and use it to construct protocols that prepare a probe state by applying the circuit and its inverse with an intervening control operation. We provide schemes that use local or global control, including a butterfly protocol with single-qubit readout. Here, the perturbation strength serves as a controlled probe, and the resulting metrological response characterizes the circuit dynamics.

We first establish a reference for the metrological response using Haar-random circuits. Their average QFI approaches its maximal value in the local protocol and grows linearly with the number of qubits under collective control. This provides a quantitative benchmark for the sensitivity generated by random-circuit dynamics. To examine whether operator spreading alone accounts for this strong response, we compare with Clifford circuits, which can spread initially local operators extensively while remaining classically simulable~\cite{mi2021information}. Their vanishing QFI shows that the benchmark probes features of the dynamics beyond operator spreading. We further develop a butterfly protocol that yields a mean inverse sensitivity growing linearly with system size, accessible through single-qubit readout. In this setting, the mean response is shared by the Haar and Clifford ensembles, while fluctuations across circuit realizations reveal their different statistics. For assessing noisy implementations, these ideal reference responses must be connected to the degradation caused by noise. Within a global white-noise model, we derive an exact relation between the noisy and ideal QFI, providing a quantitative description of how the benchmark changes with noise strength. This connects the ideal-circuit analysis to our central goal of benchmarking noisy random quantum circuits without computing ideal output probabilities.

\section*{Results}
\label{sec:Results}

\subsection*{Out-of-Time Order Correlators for Quantum Estimation}
\label{sec:Out-of-Time Order Correlation for Quantum Estimation}

To benchmark random quantum circuits through their response to controlled perturbations, we first connect this response to the underlying circuit dynamics. We quantify the response using quantum Fisher information (QFI), which determines the ultimate precision in estimating the perturbation strength~\cite{paris2009quantum}. We show that, for probe states prepared by a circuit and its inverse with an intervening operation, the QFI can be expressed in terms of out-of-time-order correlators (OTOCs)~\cite{swingle2016measuring}. This relation provides the foundation for metrological benchmarking of random quantum circuits, linking their dynamical properties to a response that can be probed without computing ideal output probabilities.

To formalize this connection, we consider an $n$-qubit circuit $\mathcal{U}$ and an initial state $\ket{0^{\otimes n}}$. We apply $\mathcal{U}$, a unitary operation $\mathcal{A}$, and the inverse circuit $\mathcal{U}^{\dagger}$ in sequence, preparing the probe state $\mathcal{A}_{\mathcal{U}}\ket{0^{\otimes n}}$, where $\mathcal{A}_{\mathcal{U}}=\mathcal{U}^{\dagger}\mathcal{A}\mathcal{U}$. To probe the circuit dependence encoded in $\mathcal{A}_{\mathcal{U}}$, we apply a perturbation $\mathcal{M}(\phi)=e^{-i\phi\mathcal{S}}$, where $\mathcal{S}$ is Hermitian, obtaining $\ket{\psi(\phi)}=e^{-i\phi\mathcal{S}}\mathcal{A}_{\mathcal{U}}\ket{0^{\otimes n}}$ (see Figs.~\ref{fig:local protocol} and~\ref{fig:global protocol}). We choose $\mathcal{S}$ such that $\mathcal{S}\ket{0^{\otimes n}}=s\ket{0^{\otimes n}}$, with $s\ne0$, so that the initial state acquires only an overall phase under $\mathcal{M}(\phi)$. A nonzero response, therefore, reflects how $\mathcal{A}_{\mathcal{U}}$ redistributes the state across different eigenspaces of $\mathcal{S}$.

To quantify this response, we use the QFI, which characterizes the sensitivity of the prepared state to changes in $\phi$ and is defined as the classical Fisher information maximized over all positive operator-valued measures (POVMs). 
For a Hermitian observable $O$, the error-propagation sensitivity $\eta_{\phi}=\Delta O/|\partial_{\phi}\langle O\rangle|$ satisfies $\eta_{\phi}\geq1/\sqrt{\mathcal{I}(\phi)}$, with the expectation value and standard deviation evaluated in $\ket{\psi(\phi)}$.
For the pure-state family above, the QFI is four times the variance of $\mathcal{S}$ in the prepared state. Using $\mathcal{S}\ket{0^{\otimes n}}=s\ket{0^{\otimes n}}$, we obtain
\begin{align}
\mathcal{I}(\phi)=\frac{4}{s^2}\left(\langle(\mathcal{S}^{\dagger})^2\mathcal{A}_{\mathcal{U}}^{\dagger}\mathcal{S}^2\mathcal{A}_{\mathcal{U}}\rangle-\langle\mathcal{S}^{\dagger}\mathcal{A}_{\mathcal{U}}^{\dagger}\mathcal{S}\mathcal{A}_{\mathcal{U}}\rangle^2\right),
\label{eq:OTOC QFI}
\end{align}
where the brackets denote expectation values in $\ket{0^{\otimes n}}$.

To make the connection with circuit dynamics explicit, we consider the OTOC $\langle\mathcal{W}^{\dagger}\mathcal{V}_{\mathcal{U}}^{\dagger}\mathcal{W}\mathcal{V}_{\mathcal{U}}\rangle$ for operators $\mathcal{W}$ and $\mathcal{V}$, where $\mathcal{V}_{\mathcal{U}}=\mathcal{U}^{\dagger}\mathcal{V}\mathcal{U}$ is the operator evolved by the circuit. In Eq.~\eqref{eq:OTOC QFI}, the first correlator corresponds to $\mathcal{V}=\mathcal{A}$ and $\mathcal{W}=\mathcal{S}^{2}$, while the second term is the square of the OTOC with $\mathcal{V}=\mathcal{A}$ and $\mathcal{W}=\mathcal{S}$. The identity thus directly relates the metrological response to OTOCs involving the circuit-evolved operator and the perturbation generator (see Supplementary Material for the derivation). Because the second OTOC enters quadratically, the ensemble-averaged QFI also probes fluctuations across circuit realizations, providing information beyond the average OTOC alone.
By using $\phi$ as a controlled parameter to probe $\mathcal{U}$, we can apply this relation to calculate reference responses for metrological benchmarking. We next develop concrete protocols using local and global control, followed by a protocol with single-qubit readout and an analysis of the effect of noise.

\subsection*{Exact Protocol}

We now use the QFI–OTOC relation to construct concrete protocols for benchmarking random quantum circuits. Following the state preparation and perturbation sequence introduced above, we consider two choices of control operations: local operations on individually addressed qubits and global operations acting across the system.

\subsubsection*{Characterization through Local Control}

\begin{figure}
    \centering
    \includegraphics[width=0.9\linewidth]{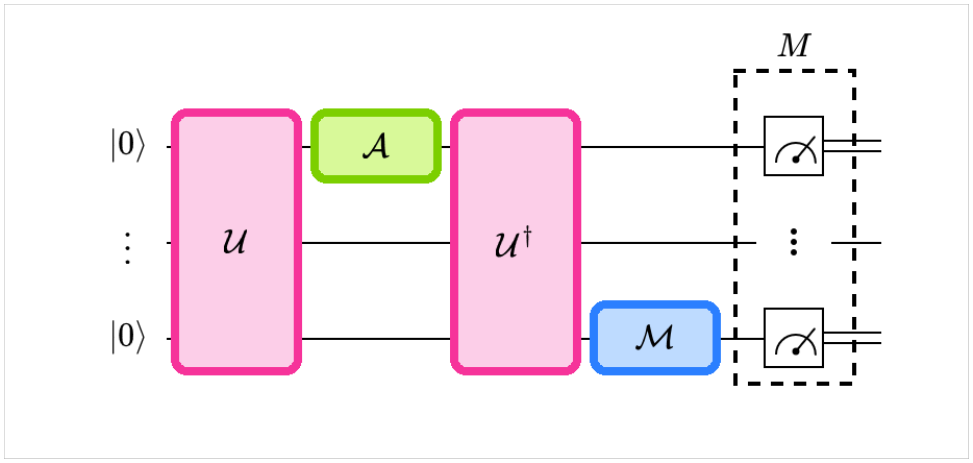}
    \caption{Schematic quantum circuit representation of the local protocol for metrological benchmarking of random quantum circuits. The protocol consists of a state preparation stage in which we apply $\mathcal{U}$ and $\mathcal{U}^{\dagger}$, interspersed with a non-identity single-qubit Pauli operation $\mathcal{A} = \mathscr{A}_{k}$ acting on qubit $k$. Then there is the application of the operation $\mathcal{M}(\phi) = m^{r}(\phi) = \left[\exp{-i \phi \mathscr{S}_{j}/2} \right]^{r}$ which consists of $r$ sequential rotations on qubit $j$, with $\mathscr{S}_{j} = \sigma^{z}_{j}$. Based on how the prepared state $ \left[ \mathcal{U}^{\dagger} \mathcal{A} \mathcal{U} \right] \ket{0^{\otimes n}} $ responds to $\mathcal{M}(\phi)$, we can characterize the unitary $\mathcal{U}$. The position of the operators $\mathcal{A}$ and $\mathcal{M}$ is only to resemble the standard structure of OTOCs; in principle, the support of the two operators may also be the same. The final measurement procedure to be adopted is discussed in both the main text and the Supplementary Material.}
    \label{fig:local protocol}
\end{figure}

We first probe the circuit dynamics through the response to a rotation on a single qubit. This provides a simple implementation of the QFI-OTOC connection, in which both the intervening operation used to prepare the probe state and the subsequent perturbation act on individual qubits, as illustrated in Fig.~\ref{fig:local protocol}. We choose $\mathcal{A}=\mathscr{A}_k$ to be a non-identity Pauli operator acting on qubit $k$, and apply the rotation $m(\phi)=e^{-i\phi\mathscr{S}_j/2}$, with $\mathscr{S}_j=\sigma_j^z$, sequentially $r$ times to qubit $j$, which may coincide with $k$~\cite{ragazzi2024generalized}. This sequence gives the perturbation $\mathcal{M}(\phi)=m^r(\phi)=e^{-ir\phi\mathscr{S}_j/2}$. Since $\mathscr{S}_j^2=\mathbb{1}$, Eq.~\eqref{eq:OTOC QFI} gives
\begin{align}
\mathcal{I}(\phi)=r^2\left(1-\langle\mathscr{S}^{\dagger}_j\mathscr{A}_{k,\mathcal{U}}^{\dagger}\mathscr{S}_j\mathscr{A}_{k,\mathcal{U}}\rangle^2\right),
\label{eq:OTOC QFI single pauli}
\end{align}
where $\mathscr{A}_{k,\mathcal{U}}=\mathcal{U}^{\dagger}\mathscr{A}_k\mathcal{U}$.

The local rotation introduces a relative phase between the two eigenspaces of $\mathscr{S}_j$, so the response depends on how the prepared state $\mathscr{A}_{k,\mathcal{U}}\ket{0^{\otimes n}}$ is distributed between them. When the two eigenspaces have equal probabilities, the state is maximally sensitive to this relative phase and the QFI reaches $r^2$. If the state is confined to either eigenspace, the rotation contributes only an overall phase, giving zero QFI.

To establish the expected metrological response in the regime of sufficiently deep random-circuit dynamics, we first consider the Haar-random ensemble as a reference, providing a natural benchmark for the QFI–OTOC relation derived above. In this limit, the QFI reads
\begin{equation}
    \label{eq:OTOC QFI Zn haar}
    \underset{\mathcal{U}\sim H}{\mathbb{E}}\left[\mathcal{I}(\phi)\right] \underset{\text{large } n}{\approx} r^2,
\end{equation}
with the corresponding finite-$n$ expression reported in the Supplementary Material. As the system size increases, the average QFI approaches its maximal value $ r^2$, indicating that Haar-random circuits prepare probes with nearly equal probabilities in the two eigenspaces of $\mathscr{S}_j$. The resulting strong response provides a reference for assessing how effectively a circuit generates sensitivity to local perturbations. 

Clifford circuits provide an informative comparison: they can spread initially local operators across the system while remaining classically simulable from the computational perspective.
This comparison allows us to isolate features of the benchmark that are not captured by operator spreading alone. In this case, $\mathscr{S}_{j}$ and $\mathscr{A}_{k,\mathcal{U}}$ commute with probability $p_{c} = \nicefrac{(2^{2n -1}-1)}{(4^{n} -1)}$ and anti-commute with probability $p_{a} = \nicefrac{(2^{2n -1})}{(4^{n} -1)}$. In either situation, $\langle \mathscr{S}_{j}^{\dagger} \mathscr{A}_{k,\mathcal{U}}^{\dagger} \mathscr{S}_{j} \mathscr{A}_{k,\mathcal{U}} \rangle^2 = 1$ and consequently, the ensemble average of Eq.~\eqref{eq:OTOC QFI single pauli} is
\begin{equation}
    \label{eq:OTOC QFI Zn clifford}
    \underset{{\mathcal{U} \sim C}}{\mathbb{E}} \left[ \mathcal{I}(\phi) \right] = 0 \, .
\end{equation}
The contrast between Eqs.~\eqref{eq:OTOC QFI Zn haar} and~\eqref{eq:OTOC QFI Zn clifford} shows that extensive operator spreading can coexist with a vanishing metrological response. The benchmark, therefore, probes features of the circuit dynamics beyond operator spreading alone. In particular, the QFI depends on the second moment of the OTOC, allowing other circuit ensembles to be assessed against the Haar reference through their response to the chosen perturbation. Intermediate values quantify the sensitivity generated by other circuit ensembles relative to the vanishing Clifford response and the near-maximal Haar reference.

\subsubsection*{Characterization through Global Control}

\begin{figure}[!ht]
    \centering
    \includegraphics[width=0.9\linewidth]{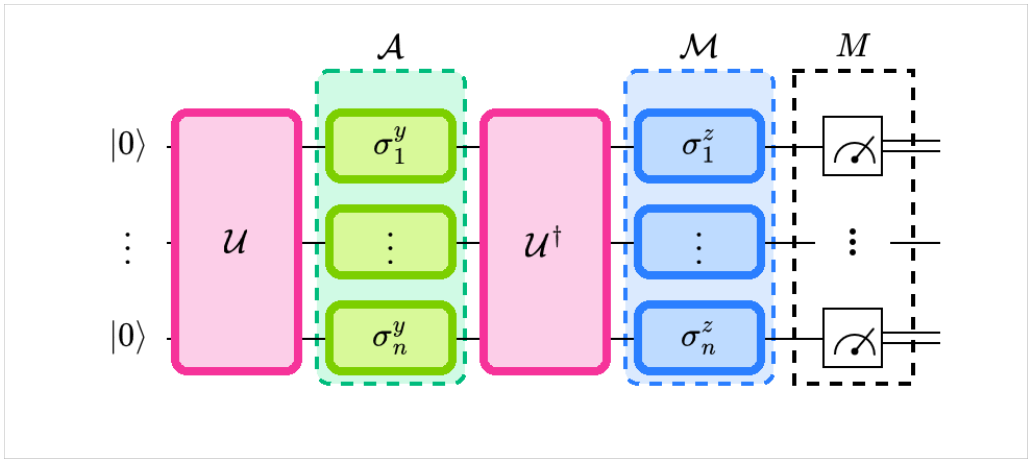}
    \caption{Schematic quantum circuit representation of the global protocol for metrological benchmarking of random quantum circuits. The protocol consists of a state preparation stage in which we apply $\mathcal{U}$ and $\mathcal{U}^{\dagger}$, interspersed with a general unitary $\mathcal{A} = \exp{-i \theta \mathcal{R}_{y}}$, where $\theta=\pi$ and $\mathcal{R}_{y} = (\nicefrac{1}{2})\sum_{j=1}^{n} \mathbb{1}^{\otimes j-1} \otimes \sigma^{y}_{j} \otimes \mathbb{1}^{\otimes n-j}$. Then there is the application of the operation $\mathcal{M}(\phi)=\left[\exp{-i\phi\mathscr{S}}\right]^{r}$, where $\mathscr{S}=(\nicefrac{1}{2})\sum_{j=1}^{n}\mathbb{1}^{\otimes j-1} \otimes \sigma^{z}_{j} \otimes \mathbb{1}^{\otimes n-j}$. Based on how the prepared state $ \left[ \mathcal{U}^{\dagger} \mathcal{A} \mathcal{U} \right] \ket{0^{\otimes n}} $ responds to $\mathcal{M}(\phi)$, we can characterize the unitary $\mathcal{U}$. The final measurement procedure to be adopted is discussed in both the main text and the Supplementary Material.}
    \label{fig:global protocol}
\end{figure}

Individual qubit addressing is not equally available across quantum platforms. In analog quantum simulators~\cite{ayral2023quantum,chakraborty2024implementing}, for example, collective rotations may be readily accessible even when selective control of individual qubits is limited. A protocol based on collective control therefore extends the benchmark to platforms with limited local addressability. In the scheme below, both the intervening preparation operation and the perturbation consist of identical single-qubit rotations applied across the system, without requiring individual qubit addressing.
In our scheme (see Fig.~\ref{fig:global protocol}), we implement a first global operation $\mathcal{A} = \exp{-i \theta \mathcal{R}_{y}}$, where $\theta=\pi$ and $\mathcal{R}_{y} = (\nicefrac{1}{2})\sum_{j=1}^{n} \mathbb{1}^{\otimes j-1} \otimes \sigma^{y}_{j} \otimes \mathbb{1}^{\otimes n-j}$, corresponding to identical rotations about the $y$ axis \footnote{A straightforward physical implementation may be the application of a magnetic field in the $y$-direction.}. $\mathcal{M}(\phi)=\left[\exp{-i\phi\mathscr{S}}\right]^{r}$ is now a collective $z$-rotation generated by the operator $\mathscr{S}=(\nicefrac{1}{2})\sum_{j=1}^{n}\mathbb{1}^{\otimes j-1} \otimes \sigma^{z}_{j} \otimes \mathbb{1}^{\otimes n-j}$.
For these control operations, Eq.~\eqref{eq:OTOC QFI} reduces to
\begin{equation}
    \label{eq:OTOC QFI Zn}
    \mathcal{I}(\phi) = 4r^2 \left(\langle\mathscr{S}^{2}\rangle-\langle\mathscr{S}\rangle^{2}\right) \, ,
\end{equation}
where here $\langle\,\cdot\,\rangle\coloneqq\bra{\chi}\,\cdot\,\ket{\chi}$ and $\ket{\chi}=\mathcal{A}_{\mathcal{U}}\ket{0^{\otimes n}}$. Again, we start with the limit of circuit-induced operators that approach the statistics of Haar-distributed unitaries. In this condition, the ensemble average of the QFI is
\begin{equation}
    \label{eq:QFI collective haar}
    \underset{{\mathcal{U} \sim H}}{\mathbb{E}}\left[\mathcal{I}(\phi)\right]\underset{\text{large } n}{\approx} n\,r^{2},
\end{equation}
where again the corresponding finite-$n$ expression is reported in the Supplementary Material. In contrast, for Clifford-random circuits, up to a global phase, $\mathcal{A}_{\mathcal{U}}=\pm P$ with $P$ a non-identity Pauli string, so $\ket{\chi}$ is, up to a global phase, a computational-basis state and therefore an eigenstate of $\mathscr{S}$. In particular, we have that $\mathscr{S} \ket{\chi} = \left( \frac{n}{2} - N_{\ket{1}} \right) \ket{\chi}$, where $N_{\ket{1}}$ represents the number of qubits in state $\ket{1}$ after the application of $\mathcal{A}_{\mathcal{U}}$ to the initial state $\ket{0^{\otimes n}}$. Independently of the value of $\left( \frac{n}{2} - N_{\ket{1}} \right)$, the state $\ket{\chi}$ is an eigenstate of $\mathscr{S}$, and consequently, the QFI is identically null for every number of qubits $n$,
\begin{equation}
    \label{eq:QFI collective clifford supp}
    \underset{{\mathcal{U} \sim C}}{\mathbb{E}}\left[\mathcal{I}(\phi)\right]=0 \, .
\end{equation}
Indeed, the present method characterizes the asymptotic random-circuit regime even when the quantum platform lacks individual qubit addressability. Thus, the local and global protocols yield different ensemble-averaged QFI for Haar-random circuits, while both give zero QFI for Clifford circuits. These two control schemes offer alternative implementations depending on the control operations available on a given quantum platform.

\paragraph*{Measurement} As with all metrology protocols, our scheme concludes with a final measurement used to infer information about the random quantum circuit. Different final measurements or methodologies may be used. Characterization of $\mathcal{U}$ may end with a standard procedure of phase estimation. Since the QFI in the Clifford-random circuit case is identically null in both global and local control characterization, we can state that any procedure aimed at estimating $\phi$ will fail independently of the final measurement applied. In contrast, when the unitary $\mathcal{U}$ resembles a Haar-random unitary, changes in $\phi$ can be detected locally using either an optimal measurement in the eigenbasis of the symmetric logarithmic derivative (SLD) or a suboptimal measurement with nonzero classical Fisher information. 

A suitable measurement choice is the projective measurement in the computational basis after applying the unitary operation $\left(\mathcal{M}(\phi_0)\mathcal{A}_{\mathcal{U}}\right)^\dagger$, with $\phi_0\approx\phi$. The probability of returning to $\ket{0^{\otimes n}}$ then reveals the response to a small change in the perturbation. For the ideal pure-state protocols, distinguishing this outcome from all other outcomes attains the QFI in the local limit $\phi\to\phi_0$, with $\phi_0$ held fixed (see Supplementary Material). Since our protocol controls the perturbation, the reference phase can be set directly, and the additional unitary employs the same circuit operations as those required for probe preparation.
Alternatively, for pure states and in the local control case, since Eq.~\eqref{eq:OTOC QFI single pauli} depends only on $\langle \mathscr{S}_{j}^{\dagger} \mathscr{A}_{k,\mathcal{U}}^{\dagger} \mathscr{S}_{j} \mathscr{A}_{k,\mathcal{U}} \rangle^2$, a suitable measurement choice to infer information about the QFI and consequently about the unitary $\mathcal{U}$ is to study the OTOC on which the QFI depends, using the standard protocols adopted in the measuring of information scrambling, from projective measurement over the initial state to interferometric procedures~\cite{swingle2016measuring} (see the Supplementary Material for details).

\subsection*{Benchmarking with Single-Qubit Readout}

The collective-control protocol makes the metrological response grow with system size, with an optimal inverse sensitivity proportional to $\sqrt{n}$ at fixed $r$ for Haar-random circuits. A butterfly sequence strengthens this scaling to a mean inverse sensitivity proportional to $n$ and makes the response accessible through a single-qubit measurement. It uses the same total-magnetization generator, combined with a modified preparation and readout sequence.
\begin{figure}[!ht]
    \centering
    \includegraphics[width=\linewidth]{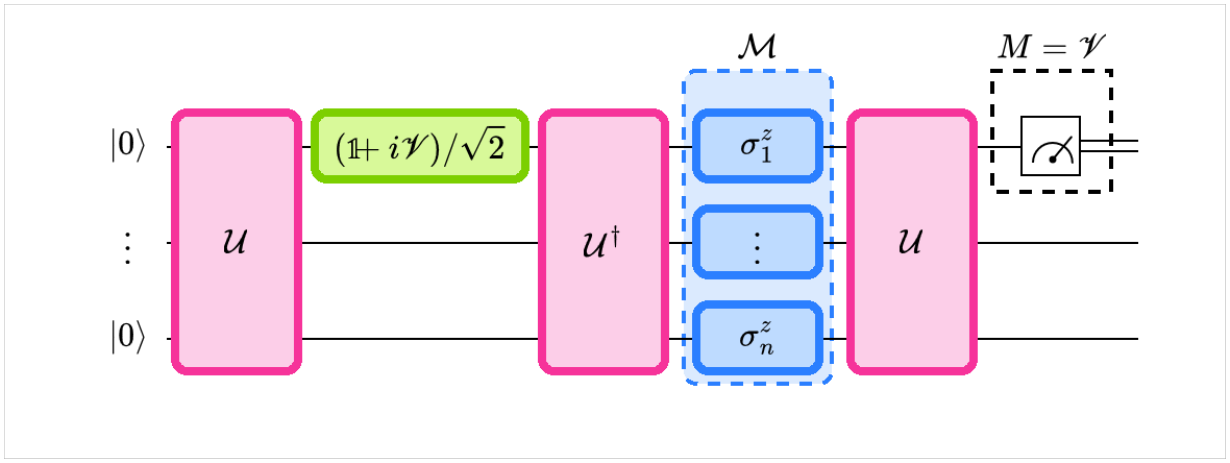}
    \caption{Schematic quantum circuit representation of the butterfly protocol for metrological benchmarking of random quantum circuits. The protocol consists of a state preparation stage in which we apply $\mathcal{U}$ and $\mathcal{U}^{\dagger}$, interspersed with a superposition of local unitaries $\nicefrac{(\mathbb{1} + i\mathscr{V})}{\sqrt{2}}$. Then there is the application of a general operation $\mathcal{M}(\phi) = \exp{(-i\phi \mathcal{S})}$, with $\mathcal{S} = \left( \nicefrac{1}{2} \right) \sum_{j=1}^{n} \mathbb{1}^{\otimes j-1} \otimes \sigma^{z}_{j} \otimes \mathbb{1}^{\otimes n-j}$. Finally, before the measurement, the unitary $\mathcal{U}$ is applied once more. The variance of the inverse sensitivity across circuit realizations distinguishes the Haar and Clifford ensembles. The detection is performed through a final local measurement $M=\mathscr{V}$.}
    \label{fig:butterfly protocol}
\end{figure}
Inspired by the butterfly protocol for quantum metrology~\cite{kobrin2024universal}, we propose a perturbative version of our protocol, according to 
\begin{equation}
    \label{eq:butterfly time evolution}
    \ket{\psi(\phi)} = \left[ \mathcal{U} \mathcal{M}(\phi) \mathcal{U}^{\dagger} \left( \frac{\mathbb{1} + i \mathscr{V}}{\sqrt{2}} \right) \mathcal{U} \right] \ket{0^{\otimes n}} \, ,
\end{equation}
where here $\mathcal{M}(\phi) = \exp{(-i\phi \mathcal{S})}$, with $\mathcal{S} = \left( \nicefrac{1}{2} \right) \sum_{j=1}^{n} \mathbb{1}^{\otimes j-1} \otimes \sigma^{z}_{j} \otimes \mathbb{1}^{\otimes n-j}$%
(see Fig.~\ref{fig:butterfly protocol}). Here, $\mathscr{V}$ is a non-identity Pauli operator acting on a single qubit, and $\mathscr{V}_{\mathcal{U}}=\mathcal{U}^{\dagger}\mathscr{V}\mathcal{U}$. Consequently, the measurement reads $M = \mathscr{V}$. The same protocol for verifying information scrambling has already been experimentally implemented~\cite{hu2026quantum}, and similar procedures are therefore practically achievable with current quantum technologies. Compared to the current literature, we will use the protocol not for the suitable creation of a useful state for quantum metrology, but for the characterization of unitaries $\mathcal{U}$ and $\mathcal{U}^{\dagger}$, analyzing both the inverse sensitivity and its variance according to a similar discrimination protocol adopted in the previous section. The final single-qubit readout is associated with a classical Fisher information that reads $F_{\mathscr{V}}^{\psi}(\phi) = \nicefrac{\left( \partial_{\phi} \langle \mathscr{V} \rangle_{\phi} \right)^2 }{\left( 1 - \langle \mathscr{V} \rangle_{\phi}^2 \right)}$. We are then interested in the expectation value of $\langle \mathscr{V} \rangle_{\phi}$, which in the weak signal regime, i.e., when $\phi \approx 0$, may be approximated as $\langle \mathscr{V} \rangle_{\phi} \approx \phi \left( \frac{n}{2} - \bra{0^{\otimes n}} \mathscr{V}_{\mathcal{U}} \mathcal{S} \mathscr{V}_{\mathcal{U}} \ket{0^{\otimes n}}  \right)$. In such conditions, and in the limit of large systems, the inverse sensitivity of the system can be approximated as \footnote{See Supplementary Material for the validity of the approximation.}
\begin{align}
    \label{eq:sensitivity weak signal regime}
    \eta^{-1}_{\phi \approx 0} & \approx \frac{n}{2} - \bra{0^{\otimes n}} \mathscr{V}_{\mathcal{U}} \mathcal{S} \mathscr{V}_{\mathcal{U}} \ket{0^{\otimes n}} \nonumber \\
    & = \frac{1}{2}  \sum_{i=1}^{n} \left(1 - \bra{0^{\otimes n}} \sigma_{i}^{z} \mathscr{V}_{\mathcal{U}} \sigma_{i}^{z} \mathscr{V}_{\mathcal{U}} \ket{0^{\otimes n}} \right) \, .
\end{align}
Then, in the weak-signal regime, the inverse sensitivity for a butterfly protocol depends only on the OTOC, not on its square. 
Averaging this response over circuit realizations yields a mean inverse sensitivity that grows linearly with the number of qubits, allowing the increased system size to enhance the response accessible through single-qubit readout.
This scaling is shared by the Haar and Clifford ensembles:
\begin{equation}
    \label{eq:ensamble average eta}
    \underset{{\mathcal{U} \sim H}}{\mathbb{E}} \left[ \eta^{-1}_{\phi \approx 0} \right] = \underset{{\mathcal{U} \sim C}}{\mathbb{E}} \left[ \eta^{-1}_{\phi \approx 0} \right] \underset{\text{large } n}{\approx} \frac{n}{2} \, .
\end{equation}
The mean response alone, therefore, does not distinguish the Haar and Clifford ensembles. To probe differences in their circuit dynamics beyond this shared mean, we examine the variance of the inverse sensitivity across circuit realizations.
Specifically, when the unitary $\mathcal{U}$ is generated by sufficiently deep random circuits, the variance of the inverse sensitivity resembles the Haar random statistics and reads as
\begin{align}
    \label{eq:variance sensitivity Haar}
     \underset{{\mathcal{U} \sim H}}{\operatorname{Var}} \left( \eta^{-1}_{\phi \approx 0} \right) \underset{\text{large } n}{\approx} \frac{n}{(4 \cdot 2^n)} \approx 0 \, .
\end{align}
This result follows from the observation that the inverse sensitivity variance is related to the OTOCs covariance matrix through $\operatorname{\operatorname{Var}}(\eta^{-1}_{\phi \approx 0}) = \frac{1}{4} \left( \sum_{i,j}^{n} \operatorname{Cov}(O_{i}, O_{j}) \right)$, where $O_{i}$ and $O_{j}$, respectively, represent the OTOCs with $\bra{0^{\otimes n}} \sigma_{i}^{z} \mathscr{V}_{\mathcal{U}} \sigma_{i}^{z} \mathscr{V}_{\mathcal{U}} \ket{0^{\otimes n}}$ and $\bra{0^{\otimes n}} \sigma_{j}^{z} \mathscr{V}_{\mathcal{U}} \sigma_{j}^{z} \mathscr{V}_{\mathcal{U}} \ket{0^{\otimes n}}$. In contrast, for Clifford random circuits, the variance of the inverse sensitivity can be written as
\begin{equation}
    \label{eq:variance sensitivity Clifford}
    \underset{{\mathcal{U} \sim C}}{\operatorname{Var}} \left( \eta^{-1}_{\phi \approx 0} \right) \underset{\text{large } n}{\approx} \frac{n}{4} \, .
\end{equation}
Equation~\eqref{eq:variance sensitivity Clifford} gives a variance that grows linearly with the number of qubits, approaching $n/4$, whereas the Haar variance in Eq.~\eqref{eq:variance sensitivity Haar} decreases as $n/(4\cdot2^n)$. 
The two ensembles, therefore, share the same mean response but exhibit increasingly different fluctuations as the system size grows.
Even when the distinction is not visible in the ensemble-averaged response, it remains encoded in circuit-to-circuit fluctuations of the inverse sensitivity, allowing the benchmark to probe circuit properties through single-qubit readout.
The butterfly protocol thus combines a response that grows with system size with access to statistical information beyond its mean. Intermediate values of the variance can be assessed against these two references to characterize the fluctuations generated by other circuit ensembles.

\section*{Benchmarking under Noise}
\label{sec:Effect of the Noise}

The results above provide reference responses for ideal random circuits.  For an experimental benchmark, however, it is equally important to understand how these responses are modified by noise. This is particularly relevant for RCS, where noise can substantially change the sampled distribution and eventually make it easier to reproduce classically. We therefore examine how the metrological response changes in a noisy implementation, taking global white noise as an analytically tractable model. This choice is motivated by studies showing that, under appropriate conditions, the effects of local noise in random circuits can be described by an effective global white-noise contribution~\cite{arute2019quantum,dalzell2024random}, with the relevant noise regimes also investigated through XEB~\cite{ware2023sharp,morvan2024phase}. Within this model, we derive an exact relation between the noisy and ideal QFI, turning the ideal responses obtained above into quantitative references for noisy implementations. This relation quantifies the degradation of the metrological benchmark without requiring the computation of ideal output probabilities.

Within this model, we describe the noisy probe by $\varrho=\mathscr{F}\ket{\psi(\phi)}\bra{\psi(\phi)}+(1-\mathscr{F})\varrho_{w.n.}$, where $\varrho_{w.n.}=\mathbb{1}/2^n$ is the maximally mixed state, and $0\leq\mathscr{F}\leq1$ is independent of $\phi$. The weight $\mathscr{F}$ determines the fidelity with the ideal probe, $\langle\psi(\phi)|\varrho|\psi(\phi)\rangle=\mathscr{F}+(1-\mathscr{F})/2^n$. The noisy SLD is proportional to the ideal SLD, so their eigenbases can be chosen to coincide. Using the relation between the SLD and QFI, we obtain
\begin{equation}
    \label{eq:noisy qfi}
    \mathcal{I}^{\varrho} (\phi) = \left( \frac{\mathscr{F}^2}{\mathscr{F} + \frac{(1-\mathscr{F})}{2^{n-1}}} \right) \mathcal{I}^{\psi} (\phi) \, .
\end{equation}
When the ideal-state weight is much larger than the white-noise weight per basis state, $2^n\mathscr{F}\gg2(1-\mathscr{F})$, this relation reduces to $\mathcal{I}^{\varrho}(\phi)\simeq\mathscr{F}\mathcal{I}^{\psi}(\phi)$. Since the white-noise weight per basis state decreases as $2^{-n}$, this regime can hold even for probe fidelities well below unity, with $\mathscr{F}$ closely approximating the probe fidelity.
Within the assumed noise model, Eq.~\eqref{eq:noisy qfi} applies to all the protocols considered above. In particular, the vanishing QFI of the Clifford circuits in the local and global protocols remains zero, whereas a nonzero ideal QFI remains nonzero for $\mathscr{F}>0$, with its magnitude reduced by the noise-dependent prefactor. For a common $\mathscr{F}$ across circuit realizations, the ensemble mean and variance of the QFI are rescaled by this prefactor and its square, respectively.

For the single-qubit readout, under the approximations considered above, we obtain $F^{\varrho}_{\mathscr{V}}(\phi) \approx \mathscr{F}^2 F^{\psi}_{\mathscr{V}}(\phi)$, with the inverse sensitivity that scales as $\eta_{\phi \approx 0}^{\varrho,-1}\approx\mathscr{F} \eta_{\phi \approx 0}^{\psi,-1}$, and its variance across circuit realizations therefore scales as $\mathscr{F}^2$. 
For a common nonzero noise strength, these relations preserve the contrast between the Haar and Clifford reference statistics while quantifying its reduction by noise. They provide quantitative references for benchmarking noisy random circuits without computing ideal output probabilities.

\section*{Discussion}
\label{sec:Discussion}

In this work, we introduced a metrological approach to benchmarking random quantum circuits through their response to controlled perturbations. The central relation between quantum Fisher information (QFI) and out-of-time-order correlators (OTOCs) connects this response to information scrambling~\cite{xu2024scrambling}. Taking the Haar ensemble as a reference for deep random circuits, we characterized the corresponding metrological response and showed that it probes statistical properties beyond those captured by the spreading of initially local operators. This is illustrated by Clifford circuits, which can exhibit extensive operator spreading while producing a qualitatively different metrological response. The benchmark can be implemented with either local or global control. We further considered a butterfly protocol with single-qubit readout, in which information beyond the ensemble-averaged response is retained in the circuit-to-circuit fluctuations.

An important aspect of RCS benchmarking is the effect of noise, which can substantially modify the sampled distribution and, in turn, make it easier to reproduce classically. For global white noise, motivated by its emergence in sufficiently random circuits~\cite{dalzell2024random}, we obtained an exact relation between the noisy and ideal QFI. The result provides a quantitative reference for how the ideal metrological response is reduced in a noisy implementation. Together, our results show that random-circuit dynamics can be benchmarked through controlled dynamical responses without requiring classical evaluation of ideal output probabilities. Extending this approach to realistic gate-dependent noise and finite-depth random circuits will be important for its application to large-scale RCS experiments.

Our results suggest several directions for further study. The theoretical relationship between QFI and OTOCs extends beyond the specific protocols analyzed here. 
In particular, this relation motivates using the susceptibility of a quantum state to probe information scrambling~\cite{hauke2016measuring}, complementing standard OTOC measurement protocols~\cite{google2025observation}.
Within the global white-noise model with a parameter-independent mixture weight, the relation between the noisy and ideal QFI applies to arbitrary pure probe states undergoing unitary parameter encoding, extending its use beyond the specific protocols considered here~\cite{ji2008parameter,meyer2021fisher}.
A natural direction is to examine how these relations change under time-dependent noise, thermalization, and device-dependent errors~\cite{sivre2019electronic,gullans2024compressed}. 
These connections between QFI, OTOCs, and noise suggest possible applications to parameterized quantum circuits~\cite{benedetti2019parameterized}, quantum metrology~\cite{meyer2021fisher}, variational quantum algorithms~\cite{cerezo2021variational}, and quantum chaos~\cite{hosur2016chaos,roberts2017chaos,balasubramanian2022quantum}, as well as related directions in quantum information~\cite{meyer2023exploiting,du2025efficient,zhou2026measuring}.

\section*{Methods}
\label{sec:Methods}

The methods adopted throughout this work are purely theoretical and based on quantum information theory and statistical methods. The relation between QFI and OTOCs is obtained from the solution of the symmetric logarithmic derivative (SLD) of pure states $\mathcal{L}^{\psi}= 2 \left( \ket{\psi(\phi)} \bra{\partial_{\phi}\psi(\phi)} + \ket{\partial_{\phi} \psi(\phi)} \bra{\psi(\phi)} \right)$ and the definition of QFI for pure states $\mathcal{I}^{\psi}(\phi) = \Tr{\ket{\psi(\phi)} \bra{\psi(\phi)} (\mathcal{L}^{\psi})^2}$. The ensemble averages of the QFI (${\mathbb{E}} \left[ \mathcal{I}(\phi) \right]$) for the exact protocols and the variance of the inverse sensitivity (${\operatorname{Var}} \left( \eta^{-1}_{\phi \approx 0} \right)$) for the perturbative case are computed using the properties of Clifford- and Haar-random circuits. In particular, the ensemble averages and variances in classically simulable circuit approximation are obtained solely through statistical calculations. In the deep random circuits scenario, the results are obtained via a symmetric decomposition of the state and verified using Haar random measurement calculations. The noise analysis, and particularly Eq.~\eqref{eq:noisy qfi}, summarizes the noise results, explicitly derived from the solution of the Lyapunov equation for SLDs $2 \partial_{\phi} \varrho =\{\varrho,\mathcal{L}^{\varrho}\}$ and the subsequent calculation of $\mathcal{I}^{\varrho}(\phi) = \Tr{\varrho (\mathcal{L}^{\varrho})^2}$. Further technical details and explicit derivations of all the results are provided in the Supplementary Material.

\begin{acknowledgments}
This work was supported by the National Research Foundation of Korea Grants (No. RS-2024-00431768 and No. RS-2025-00515456) funded by the Korean government (Ministry of Science and ICT (MSIT)) and the Institute of Information \& Communications Technology Planning \& Evaluation (IITP) Grants funded by the Korean government (MSIT) (No. RS-2024-00437284, No. IITP-2025-RS-2025-02283189 and No. IITP-2025-RS-2025-02263264) by Global Partnership Program of Leading Universities in Quantum Science and Technology (RS-2025-08542968) through the National Research Foundation of Korea~(NRF) funded by the Korean government (Ministry of Science and ICT(MSIT)). Circuits drawn with ``PsiQuantum Circuit designer": \url{https://www.psiquantum.com/news-import/open-access-circuit-designer}.
\end{acknowledgments}

\input{supplement}

\bibliography{biblio}

\end{document}

%% file: supplement.tex
\clearpage
\onecolumngrid
\setcounter{section}{0}
\setcounter{equation}{0}
\setcounter{figure}{0}
\setcounter{table}{0}
\setcounter{footnote}{0}
\setcounter{secnumdepth}{3}
\renewcommand{\thesection}{S\arabic{section}}
\renewcommand{\theequation}{S\arabic{equation}}
\renewcommand{\thefigure}{S\arabic{figure}}
\renewcommand{\thetable}{S\arabic{table}}
\renewcommand{\theHsection}{S\arabic{section}}
\renewcommand{\theHequation}{S\arabic{equation}}
\renewcommand{\theHfigure}{S\arabic{figure}}
\renewcommand{\theHtable}{S\arabic{table}}

\begin{center}
{\large\bfseries Supplementary Material for: Metrological Benchmarking of Random Quantum Circuits}\\[1ex]
{Simone Cavazzoni and Changhun Oh}
\end{center}

\par\medskip
\noindent
Here we provide all the detailed calculations required to obtain the results presented in the main text. We first provide a brief review of the theory of quantum estimation and out-of-time-order correlators (OTOCs), since the main results relate the quantum Fisher information to OTOCs for benchmarking random quantum circuits. Specifically, we start with the theoretical foundations of quantum Fisher information (QFI) and OTOCs and provide a general relation between these two quantities. We then turn to the main part of this additional material, which provides explicit calculations for local, global, and perturbative ``butterfly" protocols for scalable metrological diagnostics of random-circuit dynamics that remain experimentally accessible. Finally, we conclude this additional material with an analysis of the impact of noise on the analyzed procedure and a generalization of the procedure to testing higher $k$-designs and OTOCs.

\bigskip

\section{Quantum Fisher Information and Quantum Sensitivity}
\label{sec:Quantum Fisher Information in a Nutshell}

In this first section of the Supplementary Material, we review a fundamental quantity in quantum information theory: the quantum Fisher information (QFI). Defined as the supremum of the classical Fisher information \cite{paris2009quantum}, it defines the ultimate precision limit in the estimation of an unknown parameter $\phi$, with $\phi\in\Phi\subset {\mathbb R}$, encoded in a quantum state $\rho(\phi)$ defined on a given Hilbert space $\mathscr{H}$, where $\rho(\phi)$ is generally addressed as a quantum statistical model. The parameter is inferred from measurement outcomes through classical data processing. Any estimation procedure can be modeled as a two-step procedure, defined as follows:
\begin{enumerate}[label=(\roman*)]
    \item a positive operator-valued measure (POVM) is performed, $\left\{\Pi_x\right\}$, with $x\in X$, such that $\Pi_x \geq 0\, \forall x$, and $\sum_{x\in X} \Pi_x = {\mathbb 1}$;
    \item a data processing protocol, based on an estimator, which is a map $\hat\phi(\{x\})$ from the measurement outcome space $X\times X\dots \times X$ ($M$ times) to the domain $\Phi$, where $M$ denotes the number of repeated measurements.
\end{enumerate}
According to the postulate of quantum mechanics, the outcomes of a measurement are distributed according to the Born rule $p(x|\phi) = \hbox{Tr}\left[\rho(\phi)\, \Pi_x\right]$. Any quantum estimation strategy aims to minimize the estimator's variance, as the strategy's precision is quantified by that variance. For unbiased estimators \footnote{An estimator is said to be unbiased if  $\int_X dX\, p(x|\phi)\, \hat\phi(x) = \phi$.}, the Cram\`er-Rao theorem establishes such a bound as follows: 
\begin{equation}
    \label{eq:CRB supp}
    \operatorname{\operatorname{Var}} \, \hat \phi\geq\frac{1}{M F_{x}(\phi)} \, .
\end{equation}
This inequality provides a lower bound on the mean square error $\operatorname{\operatorname{Var}} \, \hat \phi = \mathbb{E}_{\phi}[(\hat{\phi}(\{x\})-\phi)^2]$ of any unbiased estimator of the parameter $\phi$, where $M$ is the number of measurements and with $F_{x}(\phi)$ named the classical Fisher information (FI), and defined as 
\begin{equation}
    \label{eq:FI supp}
    F_{x}(\phi)  = \sum_{x} p(x | \phi) \big[\partial_\phi \ln p(x | \phi)\big]^2\, = \sum_{x} \frac{\left( \partial_\phi p(x | \phi) \right)^2}{p(x | \phi)} \, ,
\end{equation}
where $p(x|\phi)$ is the conditional probability of obtaining the value $x$, measuring an observable X, when the parameter assumes a value $\phi$. The relations presented in Eqs.~\eqref{eq:CRB supp}-\eqref{eq:FI supp} depend on the choice of the POVM. To obtain the maximum limit on the precision of any estimation strategy for the unknown parameter $\phi$, it is therefore necessary to maximize the Fisher information over all possible POVMs. The optimal measure corresponds to the  spectral measure of the symmetric logarithmic derivative $\mathcal{L}$, which is the self-adjoint operator solving the Lyapunov equation
\begin{equation}
    \label{eq:SLD supp}
    2 \partial_\phi \rho(\phi)  = \{\mathcal{L},\rho(\phi)\}\,.
\end{equation}
The supremum of the FI is the quantum Fisher information (QFI) $\mathcal{I}(\phi)$ 
\begin{align}
    \label{eq:FI maximization supp}
    \max_{\{\Pi_x\}} F(\phi) & = \mathcal{I}(\phi) \equiv \hbox{Tr} \left[\rho(\phi)\, \mathcal{L}^2\right] \, ,
\end{align}
which, as its classical counterpart, is associated with the Quantum Cram\`er-Rao bound
\begin{align}
    \label{eq:QCRB supp}
    \operatorname{\operatorname{Var}} \, \hat \phi& \geq \frac{1}{M \mathcal{I}(\phi)} \, .
\end{align}
According to the definition of QFI, the Quantum Cramér-Rao bound provides an ultimate bound in precision for estimation problems. It depends on the geometric structure of the quantum statistical model and is measurement-independent. Thus, it can be seen as the amount of information about the parameter $\phi$ carried by the quantum system $\rho(\phi)$. For pure states $\rho(\phi)=|\psi(\phi)\rangle\langle\psi(\phi)|$ the SLD is 
\begin{equation}
    \label{eq:SLD pure state}
    \mathcal{L} = 2\left( \ket{\partial_{\phi} \psi(\phi)} \bra{\psi(\phi)} + \ket{\psi(\phi)} \bra{ \partial_{\phi} \psi(\phi)} \right) \, ,
\end{equation}
and the QFI $\mathcal{I}(\phi)$ may be rewritten as
\begin{equation}
    \label{eq:QFI pure states supp} 
    \mathcal{I}^{\psi}(\phi) = 4 \left( \abs{\partial_{\phi} \psi(\phi) }^2 + \bra{\partial_{\phi} \psi(\phi)} \ket{\psi(\phi)}^2 \right) = 4 \left( \abs{\partial_{\phi} \psi(\phi) }^2 - \abs{\bra{\partial_{\phi} \psi(\phi)} \ket{\psi(\phi)}}^2 \right) \, .
\end{equation}
The two formulations are equivalent since the term $\bra{\partial_{\phi} \psi(\phi)} \ket{\psi(\phi)}$ is purely imaginary, as $\partial_{\phi} \bra{\psi(\phi)}\ket{\psi(\phi)} = \bra{\partial_{\phi} \psi(\phi)}\ket{\psi(\phi)} + \bra{\psi(\phi)}\ket{ \partial_{\phi} \psi(\phi)} = \bra{\partial_{\phi} \psi(\phi)}\ket{\psi(\phi)} + h.c. = 2 \Re{\bra{\partial_{\phi} \psi(\phi)}\ket{\psi}} = 0$, where the last equality is valid due to norm conservation. The explicit expression of the SLD for a general non-pure state can be derived from the Lyapunov equation. Particularly, writing $\rho(\phi)$ in its eigenbasis, $ \rho(\phi) = \sum_{n} \rho_n | \lambda_n \rangle \langle \lambda_n| $ we can write $\mathcal{L}^{\rho}$ as
\begin{equation} 
	\label{eq:SDL e-basis supp}
   	\mathcal{L}^{\rho} = 2 \sum_{nm} \frac{\langle \lambda_m | \partial_\phi \rho(\phi)|\lambda_n \rangle}{\rho_n + \rho_m} | \lambda_m \rangle \langle \lambda_n | \ ,
\end{equation}
where the sum includes only terms with $\rho_n + \rho_m \neq 0$, and the eigenvalues $\rho_n$ and eigenvectors $|\lambda_n\rangle$ may depend on $\phi$. Using Eq.~\eqref{eq:SDL e-basis supp}, the quantum Fisher information $\mathcal{I}^{\rho}(\phi)$ can be rewritten as
\begin{equation}
    \mathcal{I}^{\rho}(\phi) = 2 \sum_{nm} \frac{ | \langle \lambda_m | \partial_\phi \rho(\phi)|\lambda_n \rangle |^2}{\rho_n + \rho_m} \ .
\end{equation}
Having been defined through optimization of all possible positive operator-valued measures (POVMs) of the classical Fisher information, the QFI can also be interpreted as a lower bound to sensitivity $\eta_{\phi}$, as $\eta_{\phi} \geq  \nicefrac{1}{\sqrt{\mathcal{I}(\phi)}}$, where $\eta_{\phi}$ refers to any Hermitian readout observable $O$, with $\eta_{\phi} = \nicefrac{\Delta O}{|\partial_{\phi} \langle O \rangle|}$.

\section{Out-of-Time Order Correlators}
\label{sec:A Brief Overview of Out-of-Time Order Correlations}

Another fundamental quantity in quantum information is the Out-of-Time Order Correlator (OTOC). This quantity was introduced as a measure of information scrambling and has been applied to the study of chaotic quantum dynamics \cite{swingle2016measuring}, with implications for interacting quantum systems and high-energy physics \cite{hosur2016chaos,roberts2017chaos}. A commonly used OTOC is defined as
\begin{equation}
    \label{eq:OTOC function supp}
    \mathcal{F}(t) = \mathcal{F} = \langle \mathcal{W}^{\dagger}_t \mathcal{V}^{\dagger} \mathcal{W}_t \mathcal{V} \rangle \, ,
\end{equation}
where $\mathcal{V}$ and $\mathcal{W}$ are two initially commuting unitary operators, i.e., $\left[ \mathcal{V}, \mathcal{W} \right]=0$. $\mathcal{W}_{t}$ represents the evolved operator $\mathcal{W}$ in the Heisenberg picture $\mathcal{W}_{t}=\mathcal{U}^{\dagger}(t) \mathcal{W} \mathcal{U}(t)$ and $\mathcal{U}=e^{-i\mathcal{H}t}$ is the time evolution operator of a chaotic quantum system with $\hbar=1$ for convenience \footnote{Generally speaking, the definition can be generalized to the application of any general unitary $\mathcal{U}$, even when it is not defined as a time evolution operator.}. The function $\mathcal{F}$ actually measures how quickly the system interactions cause the initially commuting operators $\mathcal{V}$ and $\mathcal{W}$ to fail to commute, 
\begin{equation}
    \label{eq:F Commutator supp}
    \langle \left[ \mathcal{W}_t, \mathcal{V} \right]^{\dagger} \left[ \mathcal{W}_t, \mathcal{V} \right] \rangle = 2 \left( 1 - \Re{\mathcal{F}} \right) \, .
\end{equation}
After the mathematical definition, we also briefly introduce the standard protocols used to measure OTOCs, which may be grouped into two different techniques. The first relies on the time evolution of a general initial state $\ket{\psi_{0}}$ according to
\begin{equation}
    \label{eq:Information Scrambling supp}
    \ket{\psi_t} = \left[ \mathcal{W}^{\dagger}_t \mathcal{V}^{\dagger} \mathcal{W}_t \mathcal{V} \right] \ket{\psi_{0}} \, ,
\end{equation}
and then performing a projective measurement $\Pi = \ket{\psi_{0}}\bra{\psi_{0}}$, since it returns the squared modulus of $\mathcal{F}$ as
\begin{equation}
    \label{eq:Modulo F supp}
    0 \leq \abs{\bra{\psi_{0}}\ket{\psi_t}}^2 = \abs{\bra{\psi_{0}} \mathcal{W}^{\dagger}_t \mathcal{V}^{\dagger} \mathcal{W}_t \mathcal{V} \ket{\psi_{0}}}^2 \leq 1 \,. 
\end{equation}
The main limitation of this approach is that it provides only the squared modulus of $\mathcal{F}$ and does not specify its components. To gain information beyond the squared modulus of $\mathcal{F}$, we have to rely on a second approach, which is based on an interferometric measurement of the state
\begin{equation}
    \label{eq:Interferometric Information Scrambling supp}
    \ket{\Psi} = \frac{\left( \mathcal{V} \mathcal{W}_{t} \ket{\psi_{0}} \right) \ket{0}_{c} + \left( \mathcal{W}_{t} \mathcal{V} \ket{\psi_{0}}  \right) \ket{1}_{c} }{\sqrt{2}} \, ,
\end{equation}
where $\{\ket{0}_{c},\ket{1}_{c}\}$ represents a control qubit. Measurement of the control qubit on the $x$-basis $\{ \ket{+}_{x} , \ket{-}_{x} \}$ with $\ket{+}_{x} = \nicefrac{\left( \ket{0} + \ket{1} \right)}{\sqrt{2}}$ and $\ket{-}_{x} = \nicefrac{\left( \ket{0} - \ket{1} \right)}{\sqrt{2}}$ we have 
\begin{equation}
    \label{eq:probability x basis supp}
    \begin{cases}
        P(\ket{+}_{x}) = \frac{1}{2} \left( 1 + \Re{\langle \mathcal{W}^{\dagger}_t \mathcal{V}^{\dagger} \mathcal{W}_t \mathcal{V} \rangle} \right) \\
        P(\ket{-}_{x}) = \frac{1}{2} \left( 1 - \Re{\langle \mathcal{W}^{\dagger}_t \mathcal{V}^{\dagger} \mathcal{W}_t \mathcal{V} \rangle} \right)
    \end{cases} \, .
\end{equation}
Complementarily, measuring the control qubit in the $y$-basis $\{ \ket{+}_{y} , \ket{-}_{y} \}$ with $\ket{+}_{y} = \nicefrac{\left( \ket{0} + i \ket{1} \right)}{\sqrt{2}}$ and $\ket{-}_{y} = \nicefrac{\left( \ket{0} - i \ket{1} \right)}{\sqrt{2}}$ we have 
\begin{equation}
    \label{eq:probability y basis supp}
    \begin{cases}
        P(\ket{+}_{y}) = \frac{1}{2} \left( 1 + \Im{\langle \mathcal{W}^{\dagger}_t \mathcal{V}^{\dagger} \mathcal{W}_t \mathcal{V} \rangle} \right) \\
        P(\ket{-}_{y}) = \frac{1}{2} \left( 1 - \Im{\langle \mathcal{W}^{\dagger}_t \mathcal{V}^{\dagger} \mathcal{W}_t \mathcal{V} \rangle} \right)
    \end{cases} \, .
\end{equation}
Measurement of the expectation values $\langle X \rangle_{c}$ and $\langle Y \rangle_{c}$, respectively, corresponding to the measurement through the operators $\{ \ket{+}_{x} {}_{x}\bra{+}$, $\ket{-}_{x} {}_{x}\bra{-} \}$ and $\{ \ket{+}_{y} {}_{y}\bra{+}$, $\ket{-}_{y} {}_{y}\bra{-} \}$ on the basis of the control qubit, leads to quantitative knowledge of $\mathcal{F}$ as a whole
\begin{equation}
    \label{eq:Interferometric F supp}
    \mathcal{F} = \langle \mathcal{W}^{\dagger}_t \mathcal{V}^{\dagger} \mathcal{W}_t \mathcal{V} \rangle = \langle X \rangle_{c} + i \langle Y \rangle_{c} \, .
\end{equation}
According to the same method, we can obtain information about $\mathcal{F}$ by looking at the purity of the reduced density matrix from an initial state in the form of Eq.~\eqref{eq:Interferometric Information Scrambling supp}, as
\begin{equation}
    \label{eq:reduced density state supp}
    \varrho_{s} = \Tr_{c} \{ {\ket{\Psi} \bra{\Psi}} \} = \frac{1}{2} \left(  \mathcal{V} \mathcal{W}_{t} \ket{\psi_{0}} \bra{\psi_{0}} \mathcal{W}_{t}^{\dagger} \mathcal{V}^{\dagger} + \mathcal{W}_{t} \mathcal{V} \ket{\psi_{0}} \bra{\psi_{0}} \mathcal{V}^{\dagger} \mathcal{W}_{t}^{\dagger} \right) \, ,
\end{equation}
where the trace $\Tr_{c}{\{\cdot\}}$ is the partial trace over the control qubit c. Additionally, $\Tr{\varrho_{s}^{2}} = \nicefrac{\left( 1 + \abs{\mathcal{F}^2} \right)}{2}$ and so on the two-dimensional branch span, when $\Tr{\varrho_{s}^{2}} = 1$, the OTOC has unit modulus, i.e., $|\mathcal{F}|=1$. In the same way, when the OTOC is null, i.e., $\mathcal{F}=0$, the reduced state is maximally mixed, and $\Tr{\varrho_{s}^{2}} = \nicefrac{1}{2}$. OTOCs may be found in a similar, but not equal, structure compared to Eq.~\eqref{eq:OTOC function supp}. Indeed, it is possible to consider the scrambling of information according to the evolution in time of the operator $\mathcal{V}$. This leads to an OTOC in the form $\langle \mathcal{W}^{\dagger} \mathcal{V}^{\dagger}_{t} \mathcal{W} \mathcal{V}_{t} \rangle$. Throughout the main text and this Supplementary Material, we consider the scrambling of information according to a general unitary $\mathcal{U}$, not necessarily connected to a time evolution, and we will refer to OTOCs in the form $\mathcal{F} = \langle \mathcal{W}^{\dagger} \mathcal{V}^{\dagger}_{\mathcal{U}} \mathcal{W} \mathcal{V}_{\mathcal{U}} \rangle$.

\section{Out-of-Time Order Correlators for Quantum Estimation}
\label{supp:sec:Out-of-Time Order Correlation for Quantum Estimation}

In this third section of the additional material, we derive a relation between QFI and OTOCs for pure-state probes. In the main text, we consider the input state $\ket{\psi_{0}} = \ket{0^{\otimes n}}$ \footnote{The results of the QFI are the same for the perturbative case defined in the main text $\ket{\psi(\phi)} = \left[ \mathcal{U} \mathcal{M}(\phi) \mathcal{U}^{\dagger} \left( \frac{\mathbb{1} + i\mathscr{V}}{\sqrt{2}} \right) \mathcal{U} \right] \ket{0^{\otimes n}}$. Explicit calculations can be found in Sec.~\ref{sec:Additional Remarks}. }, which consists of the standard input state considered in any quantum circuit setup. Then we apply a sequence of operations in the form
\begin{equation}
    \label{eq:evolution initial state supp}
    \ket{\psi(\phi)} = \mathcal{M}(\phi) \mathcal{U}^{\dagger} \mathcal{A} \mathcal{U} \ket{\psi_{0}} \, ,
\end{equation}
where $\mathcal{M}(\phi) = \exp{-i\phi \mathcal{S}}$, with $\mathcal{S} = \mathcal{S}^{\dagger}$ Hermitian and $\mathcal{S} \ket{0^{\otimes n}} = s \ket{0^{\otimes n}}$, with $s\neq0$. According to Eq.~\eqref{eq:QFI pure states supp}, the QFI of $\phi$ for the pure state $\ket{\psi(\phi)}$ may be rewritten as
\begin{equation}
    \label{eq:QFI pure state protocol supp}
    \mathcal{I}(\phi) = 4 \left( \langle \mathcal{A}_{\mathcal{U}}^{\dagger} \mathcal{S}^2 \mathcal{A}_{\mathcal{U}} \rangle - \langle \mathcal{A}_{\mathcal{U}}^{\dagger} \mathcal{S} \mathcal{A}_{\mathcal{U}} \rangle^2 \right) = \frac{4}{s^2} \left( \langle (\mathcal{S}^{\dagger})^2 \mathcal{A}_{\mathcal{U}}^{\dagger} \mathcal{S}^2 \mathcal{A}_{\mathcal{U}} \rangle - \langle \mathcal{S}^{\dagger} \mathcal{A}_{\mathcal{U}}^{\dagger} \mathcal{S} \mathcal{A}_{\mathcal{U}} \rangle^2 \right) \, ,
\end{equation}
where $\mathcal{A}_{\mathcal{U}} = \mathcal{U}^{\dagger} \mathcal{A} \mathcal{U}$ and the brackets denote expectation values in $\ket{0^{\otimes n}}$.  The purpose of the protocol is to propose an experimentally accessible and scalable metrological diagnostic of random-circuit dynamics, by probing the sensitivity of the resulting state, quantified by the QFI, to a controlled operation $\mathcal{M}(\phi)$. In standard metrological procedures, the goal is to estimate $\phi$; here, $\mathcal{M}(\phi)$ serves as a controlled probe of random-circuit dynamics. Eq.~\eqref{eq:QFI pure state protocol supp} is essentially the difference between OTOCs and provides a relation between the quantum Fisher information and the standard measurement of information scrambling and randomness. As we demonstrate in the main text, this connection allows us to distinguish among different random circuit dynamics, from scrambling circuits that remain classically efficient to simulate to the limit of sufficiently deep random circuits that fully resemble the Haar-random statistics. The 
formula presented in Eq.~\eqref{eq:QFI pure state protocol supp} can be generalized to any isolated pure quantum state $\ket{\psi_{0}}$ for any operation in the form $\ket{\psi(\phi)} = \exp{-i\phi H} \ket{\psi_{0}}$. Indeed, any state $\ket{\psi_{0}}$ can be written as $\ket{\psi_{0}} = P \ket{h}$, where $P$ is unitary and $H\ket{h}=h\ket{h}$ with $h\neq0$. With the expectation values now evaluated in $\ket{h}$, the QFI takes the form
\begin{equation}
    \label{eq:QFI phase estimation supp}
    \mathcal{I}(\phi) = 4 \left( \langle P^{\dagger} H^{2} P \rangle - \langle P^{\dagger} H P  \rangle^2 \right) = \frac{4}{h^2} \left( \langle (H^{\dagger})^2 P^{\dagger} H^2 P  \rangle - \langle H^{\dagger} P^{\dagger} H P  \rangle^2 \right) \, .
\end{equation}
Additionally, the formula presented in Eqs.~\eqref{eq:QFI pure state protocol supp},\eqref{eq:QFI phase estimation supp} can be generalized to any isolated pure quantum state $\ket{\psi_{0}}$ subject to a unitary evolution. Indeed, we can consider any parameter-dependent state $\ket{\psi(\phi)}$ as the result of a parameter-dependent unitary operator $\mathcal{O}(\phi)$ \footnote{Operator $\mathcal{O}(\phi)$ may include both the state preparation and the actual operation that encode the parameter in the state as $\mathcal{O}(\phi) = \Tilde{\mathcal{O}}(\phi) \mathcal{P}$, with $\mathcal{P}$ state preparation operation and $\Tilde{\mathcal{O}}(\phi)$ parameter dependent operator.} applied on an initial parameter-independent state $\ket{\psi_{0}}$. The state is then $\ket{\psi(\phi)} = \mathcal{O}(\phi) \ket{\psi_{0}}$ and its derivative may be written as $\ket{\partial \psi(\phi)} = \left( \partial \mathcal{O}(\phi) \right) \ket{\psi_{0}} = \left( \partial \mathcal{O}(\phi) \right) \mathcal{O}^{\dagger}(\phi) \ket{\psi(\phi)}$. The QFI can be related to the variance of a Hermitian operator \cite{liu2015quantum} in the form $K = i \left( \partial \mathcal{O}^{\dagger}(\phi)\right) \mathcal{O}(\phi)$, as
\begin{align}
    \label{eq:general QFI supp}
    \mathcal{I}(\phi) &= 4 \left( \operatorname{\operatorname{Var}} \left( K \right)_{\ket{\psi_{0}}} \right) \, .
\end{align}
The trivial case $K=0$ has zero QFI. For $K\neq0$, we can reconstruct the out-of-time-order structure by rewriting Eq.~\eqref{eq:general QFI supp} as
\begin{align}
    \label{eq:general QFI OTOC supp}
    \mathcal{I}(\phi) &=  4 \left( \bra{\psi_{0}} K^2  \ket{\psi_{0}} - \bra{\psi_{0}} K \ket{\psi_{0}}^2 \right) = 4 \left( \bra{\psi_{0}} V^{\dagger} V K^2  V^{\dagger} V  \ket{\psi_{0}} - \bra{\psi_{0}}  V^{\dagger} V K  V^{\dagger} V \ket{\psi_{0}}^2 \right) \nonumber \\
    &= 4 \bigg\{ \bra{k} V K^2  V^{\dagger} \ket{k} - \bra{k} V K  V^{\dagger} \ket{k}^2 \bigg\} = \frac{4}{k^2} \bigg\{ \bra{k} (K^{\dagger})^2 V K^2  V^{\dagger} \ket{k} - \bra{k} K^{\dagger} V K  V^{\dagger} \ket{k}^2 \bigg\} \, ,
\end{align}
where $V$ is unitary, $V\ket{\psi_{0}}=\ket{k}$, and $K\ket{k}=k\ket{k}$. Equation~\eqref{eq:general QFI OTOC supp} requires $k\neq0$, which can always be chosen for $K\neq0$. The analogous representations in Eqs.~\eqref{eq:QFI pure state protocol supp} and~\eqref{eq:QFI phase estimation supp} require $s\neq0$ and $h\neq0$, respectively. With $\mathcal{V}=V^{\dagger}$, this gives an algebraic representation of the pure-state QFI in terms of correlators with an out-of-time-order structure. 

\section{Calculation Details}
\label{sec:Calculation Details}

In this section, we explicitly report all the detailed calculations related to our protocol, focusing on a system of $n$ qubits, with Hilbert space $\mathscr{H} \cong \mathbb{C}^{d}$, $d = 2^{n}$. Since the aim of our work is to provide a simple and scalable metrological diagnostic of random circuits, we have to rely on ensemble average calculations to verify the nature of unitary $\mathcal{U}$. As stated also in the main text, as a case study, we report two explicit limiting cases: (i) strongly scrambling circuits that remain classically efficient to simulate, i.e., Clifford-random circuits, and (ii) a limit of sufficiently deep random circuits that fully resemble the Haar-random statistics. Since the Haar- and Clifford-random ensemble averages are equal when three or fewer copies of $\mathcal{U}$ and $\mathcal{U}^{\dagger}$ are involved \cite{webb2015clifford,zhu2016clifford}, our protocol is based on ensemble averages that involve $4$ copies of the unitary $\mathcal{U}$ and $\mathcal{U}^{\dagger}$ (see Eq.~\eqref{eq:QFI pure state protocol supp}). In the following section, we present all calculations needed to obtain the results in the main text for the exact and perturbative protocols.

\subsection{Exact Protocol}

The exact protocols presented in the main text are based on operations with global or local control. Both protocols can be summarized as $\ket{\psi(\phi)} = \left[  \mathcal{M}(\phi) \mathcal{A}_{\mathcal{U}} \right] \ket{0^{\otimes n}}$, where $\mathcal{A}_{\mathcal{U}} = \mathcal{U}^{\dagger} \mathcal{A} \mathcal{U}$, where, depending on the support of the operators $\mathcal{A}$ and $\mathcal{M}(\phi)$, the protocols are referred to as local or global.

\subsubsection*{Characterization through Local Control}

We then consider the operations $\mathcal{A} = \mathscr{A}_{k}$ to be a single qubit Pauli operation different from the identity that acts on the qubit $k$, and $\mathcal{M}(\phi) = m^{r}(\phi) = \left[\exp{-i \phi \mathscr{S}_{j}/2} \right]^{r}$ to be the sequential application of rotations $\mathscr{S}_{j} = \sigma^{z}_{j}$ on the qubit $j$ \cite{ragazzi2024generalized}. In this scenario, Eq.~\eqref{eq:QFI pure state protocol supp} is reduced to
\begin{equation}
    \label{eq:OTOC QFI single pauli supp}
    \mathcal{I}(\phi) = r^2 \left( 1 - \langle \mathscr{S}_{j}^{\dagger} \mathscr{A}_{k,\mathcal{U}}^{\dagger} \mathscr{S}_{j} \mathscr{A}_{k,\mathcal{U}} \rangle^2 \right) \, ,
\end{equation}
where the OTOC $\langle \mathscr{S}_{j}^{\dagger} \mathscr{A}_{k,\mathcal{U}}^{\dagger} \mathscr{S}_{j} \mathscr{A}_{k,\mathcal{U}} \rangle$ is purely real as $\langle \mathscr{S}_{j}^{\dagger} \mathscr{A}_{k,\mathcal{U}}^{\dagger} \mathscr{S}_{j} \mathscr{A}_{k,\mathcal{U}} \rangle = \langle \mathscr{S}_{j}^{\dagger} \mathscr{A}_{k,\mathcal{U}}^{\dagger} \mathscr{S}_{j} \mathscr{A}_{k,\mathcal{U}} \rangle^{*}$. To establish the expected metrological response in the regime of sufficiently deep random-circuit dynamics, we first consider the Haar-random ensemble as a reference, providing a natural benchmark for the QFI–OTOC relation derived above. To explicitly calculate OTOC in this limit, we rely on a symmetry decomposition of the state $\ket{\chi}=\mathscr{A}_{k,\mathcal{U}} \ket{0^{\otimes n}}$. In particular, OTOC $\langle \mathscr{S}_{j}^{\dagger} \mathscr{A}_{k,\mathcal{U}}^{\dagger} \mathscr{S}_{j} \mathscr{A}_{k,\mathcal{U}} \rangle$, since $\bra{0^{\otimes n}} \mathscr{S}_{j}^{\dagger} = \bra{0^{\otimes n}}$ can be rewritten as $\bra{\chi} \mathscr{S}_{j} \ket{\chi}$. If we denote the $(d-1)$-dimensional subspace orthogonal to $\ket{0^{\otimes n}}$ as $\mathscr{H}_{0^{\otimes n}}^\perp$ we can write 
\begin{equation}
    \label{eq:decomp supp}
    \ket{\chi}=X\ket{0^{\otimes n}}+\sqrt{1-X^{2}}\,\ket{\mu},\quad
    \ket{\mu}\in\mathscr{H}_{0^{\otimes n}}^\perp,
\end{equation}
with $X\coloneqq\braket{0^{\otimes n}}{\chi}$, overlap of $\ket{\chi}$ with the initial state $\ket{0^{\otimes n}}$, and $\ket{\mu}$ unit vector orthogonal to $\ket{0^{\otimes n}}$. Under the Haar measure, the scalar $X$ and the direction $\ket{\mu}$ are statistically independent, and $\ket{\mu}$ is uniformly distributed on the unit sphere of $\mathbb{C}^{d-1}$. The subspace $\mathscr{H}_{0^{\otimes n}}^\perp$ is invariant under $\mathscr{S}_{j}$, as $\mathscr{S}_{j} \ket{0^{\otimes n}} = \ket{0^{\otimes n}}$, and then for any $\ket{\nu}\perp\ket{0^{\otimes n}}$, $\bra{0^{\otimes n}} \mathscr{S}_{j} \ket{\nu}=\bra{0^{\otimes n}}\ket{\nu}=0$. Then, by denoting $\Tilde{\mathscr{S}_{j}}$ the restriction of $\mathscr{S}_{j}$ to $\mathscr{H}_{0^{\otimes n}}^\perp$, and due to the identities $\mathscr{S}_{j}^{2} = \mathbb{1}$ and $\Tr{\mathscr{S}_{j}} = 0$, we have $\Tilde{\mathscr{S}_{j}}^2 = \mathbb{1}_{d-1}$ and $\Tr{\Tilde{\mathscr{S}_{j}}} = \Tr{\mathscr{S}_{j}} - \bra{0^{\otimes n}}  \mathscr{S}_{j} \ket{0^{\otimes n}} = -1$. Finally, according to the decomposition in Eq.~\eqref{eq:decomp supp}, we can write $\langle \mathscr{S}_{j}^{\dagger} \mathscr{A}_{k,\mathcal{U}}^{\dagger} \mathscr{S}_{j} \mathscr{A}_{k,\mathcal{U}} \rangle = X^{2}+(1-X^{2})\,Y$, with $Y\coloneqq\bra{\mu}\Tilde{\mathscr{S}}_{j}\ket{\mu}$. After decomposing the state, we can compute the ensemble average of the OTOC. First, we recall that for a general Haar-random state $\ket{\zeta}\in\mathbb{C}^{d'}$ and for a general operator $M$ we have the relations 
\begin{align}
    \label{eq:Haar random state E averages supp}
    \underset{{\zeta \sim H}}{\mathbb{E}}\bigl[\bra{\zeta}M\ket{\zeta}\bigr] &=\frac{\Tr M}{d'}, \nonumber \\
    \underset{{\zeta \sim H}}{\mathbb{E}}\bigl[\bra{\zeta}M\ket{\zeta}^{2}\bigr] &=\frac{(\Tr M)^{2}+\Tr(M^{2})}{d'(d'+1)},  \\
    \underset{{\zeta \sim H}}{\mathbb{E}}[\bra{\zeta}M\ket{\zeta}^{4}] &= \frac{(\Tr M)^{4}+6\Tr(M^{2}) (\Tr M)^2 + 3 (\Tr M^{2})^2 + 8 (\Tr M^3)\Tr(M) + 6 \Tr(M^4)}{d'(d'+1)(d'+2)(d'+3)} \nonumber \, .
\end{align}
When the operator $M$ is a nonidentity Pauli string $\Tr(M^{2k})=d^{'}$, $\Tr(M^{2k+1})=0$ and then we can explicitly calculate the ensemble averages of $X^2$ and $X^4$ according to the formulas presented in Eq.~\eqref{eq:Haar random state E averages supp}. In particular, since for $X$ $d^{'} = d$, $\underset{{\zeta \sim H}}{\mathbb{E}}[X^2] = \frac{1}{d+1}$ and $\underset{{\zeta \sim H}}{\mathbb{E}}[X^4] = \frac{3}{(d+1)(d+3)}$, we have
\begin{align}
    \label{eq:aux1 supp}
    \underset{{\zeta \sim H}}{\mathbb{E}}[X^2(1-X^2)]=\frac{d}{(d+1)(d+3)} \, , \, \, \underset{{\zeta \sim H}}{\mathbb{E}}[(1-X^2)^2]  =\frac{d(d+2)}{(d+1)(d+3)} \, .
\end{align}
Analogously for the ensemble averages of $Y$, we have $d'=d-1$, and then 
\begin{align}
    \label{eq:EY supp}
    \underset{{\zeta \sim H}}{\mathbb{E}} [Y] &= \frac{-1}{d-1} \, , \, \, \underset{{\zeta \sim H}}{\mathbb{E}} [Y^2] = \frac{1}{d-1} \, .
\end{align}
Having all the ensemble averages of $X$ and $Y$ we can compute the OTOC. Due to the independence of $X$ and $Y$ (related to the decomposition presented in Eq.~\eqref{eq:decomp supp}) we can rewrite $\underset{{\mathcal{U} \sim H}}{\mathbb{E}} \left[ \langle \mathscr{S}_{j}^{\dagger} \mathscr{A}_{k,\mathcal{U}}^{\dagger} \mathscr{S}_{j} \mathscr{A}_{k,\mathcal{U}} \rangle \right]$ as 
\begin{align}
    \label{eq:EC supp}
    \underset{{\mathcal{U} \sim H}}{\mathbb{E}} \left[ \langle \mathscr{S}_{j}^{\dagger} \mathscr{A}_{k,\mathcal{U}}^{\dagger} \mathscr{S}_{j} \mathscr{A}_{k,\mathcal{U}} \rangle \right] =\underset{{\zeta \sim H}}{\mathbb{E}} [X^2]+\underset{{\zeta \sim H}}{\mathbb{E}} [1-X^2]\,\underset{{\zeta \sim H}}{\mathbb{E}} [Y] =\frac{1}{d+1}-\frac{d}{d+1}\cdot\frac{1}{d-1} =-\frac{1}{d^2-1}.
\end{align}
The ensemble average of the OTOC squared can be calculated as
\begin{align}
    \label{eq:EC2 supp}
    \underset{{\mathcal{U} \sim H}}{\mathbb{E}} \left[ \langle \mathscr{S}_{j}^{\dagger} \mathscr{A}_{k,\mathcal{U}}^{\dagger} \mathscr{S}_{j} \mathscr{A}_{k,\mathcal{U}} \rangle^2 \right] &=\underset{{\zeta \sim H}}{\mathbb{E}}[X^4] +2\,\underset{{\zeta \sim H}}{\mathbb{E}}[X^2(1\!-\!X^2)]\,\underset{{\zeta \sim H}}{\mathbb{E}}[Y] +\underset{{\zeta \sim H}}{\mathbb{E}} [(1\!-\!X^2)^2]\, \underset{{\zeta \sim H}}{\mathbb{E}}[Y^2]\notag\\
    &=\frac{3}{(d+1)(d+3)} -\frac{2d}{(d-1)(d+1)(d+3)} +\frac{d(d+2)}{(d-1)(d+1)(d+3)}\notag\\
    &=\dfrac{d^2+3d-3}{(d-1)(d+1)(d+3)}.
\end{align}
According to these results, when the unitaries $\mathcal{U}$ and $\mathcal{U}^{\dagger}$ resemble the Haar-random statistics, the ensemble average of the QFI is
\begin{equation}
    \label{eq:OTOC QFI Zn haar supp}
    \underset{{\mathcal{U} \sim H}}{\mathbb{E}} \left[ \mathcal{I}(\phi) \right] = r^2 \left( 1 - \frac{d^2+3d-3}{(d-1)(d+1)(d+3)} \right) \, ,
\end{equation}
where $d=2^n$ is the dimensionality of the Hilbert space. We then compare this result with the Clifford-random circuit benchmark to isolate features of the benchmark that are not captured by operator spreading alone. In this limit, the operator $\mathcal{U}^{\dagger} \mathscr{A}_{k} \mathcal{U}$ is a Pauli string up to a sign, since the unitary $\mathcal{U}$ is a Clifford-random unitary and Clifford operators map Pauli operators onto Pauli operators under conjugation $\mathscr{A}_{k,\mathcal{U}} = \mathcal{U}^{\dagger} \mathscr{A}_{k} \mathcal{U} \in \mathcal{P}_n\setminus\{\mathbb{1}\}$. Additionally, when $\mathcal{U}$ is random, $\mathscr{A}_{k,\mathcal{U}}$ is uniformly distributed over all the $4^n-1$ non-identity Pauli strings. The operator $\mathscr{S}_{j}$ is a single Pauli string that has support on a single qubit of our system, and commutes with $2^{2n-1} -1$ Pauli strings (i.e., half of the Pauli strings minus the identity); and anti-commutes with $2^{2n-1}$ Pauli strings. Then, operators $\mathscr{S}_{j}$ and $\mathscr{A}_{k,\mathcal{U}}$ commute with probability $p_{c} = \nicefrac{(2^{2n -1}-1)}{4^{n} -1}$ and anti-commute with probability $p_a = \nicefrac{(2^{2n -1})}{4^{n} -1}$. The ensemble average of the OTOC is then 
\begin{equation}
    \label{eq:OTOC clifford ensemble average supp}
    \underset{{\mathcal{U} \sim C}}{\mathbb{E}} \left[ \langle \mathscr{S}_{j}^{\dagger} \mathscr{A}_{k,\mathcal{U}}^{\dagger} \mathscr{S}_{j} \mathscr{A}_{k,\mathcal{U}} \rangle \right] = p_c - p_a = -\frac{1}{d^{2}-1} \, .
\end{equation}
This result reproduces exactly the results obtained in Eq.~\eqref{eq:EC supp}. This derives from the fact that Clifford-random unitaries reproduce the Haar-random results for every expectation value that involves $2$ copies of $\mathcal{U}$ and $\mathcal{U}^{\dagger}$, and intrinsically verify the validity of the adopted symmetric decomposition protocol. To distinguish the two ensembles, we calculate $\underset{{\mathcal{U} \sim C}}{\mathbb{E}} \left[ \langle \mathscr{S}_{j}^{\dagger} \mathscr{A}_{k,\mathcal{U}}^{\dagger} \mathscr{S}_{j} \mathscr{A}_{k,\mathcal{U}} \rangle^2 \right]$. Because of the (anti-)commutation relation of Pauli strings, $\langle \mathscr{S}_{j}^{\dagger} \mathscr{A}_{k,\mathcal{U}}^{\dagger} \mathscr{S}_{j} \mathscr{A}_{k,\mathcal{U}} \rangle^2 = 1$ and consequently, the ensemble average for Clifford-random circuits is $\underset{{\mathcal{U} \sim C}}{\mathbb{E}} \left[ \langle \mathscr{S}_{j}^{\dagger} \mathscr{A}_{k,\mathcal{U}}^{\dagger} \mathscr{S}_{j} \mathscr{A}_{k,\mathcal{U}} \rangle^2 \right] = +1$ and finally, according to Eq.~\eqref{eq:OTOC QFI single pauli supp}
\begin{equation}
    \label{eq:OTOC QFI Zn clifford supp}
    \underset{{\mathcal{U} \sim C}}{\mathbb{E}} \left[ \mathcal{I}(\phi) \right] = 0 \, .
\end{equation}
This means that independently of the number of qubits, the QFI is identically null. For Clifford-random circuits, the ensemble average is then identically zero regardless of the Hilbert space dimensionality; therefore, the present protocol is an experimentally accessible and scalable metrological diagnostic of random-circuit dynamics. Conversely, for Haar-random statistics, the QFI results depend on the Hilbert space dimension, highlighting a clear difference between the two cases. More interestingly, the difference is becoming more and more evident as the dimensionality of the Hilbert space increases, as in the limit of a high number of qubits, Eq.~\eqref{eq:OTOC QFI Zn haar supp} scales as $\underset{{\mathcal{U} \sim H}}{\mathbb{E}} \left[ \mathcal{I}(\phi) \right] \approx r^2 \left( 1 - \frac{1}{d} \right) \approx r^2$. This large-system limit provides a near-maximal Haar reference against which the metrological response of other circuit ensembles can be assessed. According to the theory of information scrambling presented in Sec.~\ref{sec:A Brief Overview of Out-of-Time Order Correlations}, the two results presented correspond to the two limits of OTOC values, providing direct insights into information scrambling through the state’s susceptibility. Intermediate results between Eq.~\eqref{eq:OTOC QFI Zn clifford supp} and Eq.~\eqref{eq:OTOC QFI Zn haar supp} can be interpreted as scenarios in which the quantum circuit is partially able to reproduce the statistics of Haar randomness.

\subsubsection*{Characterization through Global Control}

After defining the protocol, in which qubits may be assessed individually, we propose a scheme in which only global operations on all qubits are allowed, assuming access to collective rotations. The derivation presented in the previous section can be generalized to the case in which $\mathscr{S}_{j}$ or $\mathscr{A}_{k}$ acts on all the qubits in the system in a translationally invariant way. In this scenario, we consider $\mathcal{A}$ to be an arbitrary non-identity Pauli string (e.g.\ $\mathcal{A}=(\sigma^y)^{\otimes n}$, with the properties $\Tr(\mathcal{A}^{2k})=d$ and $\Tr(\mathcal{A}^{2k+1})=0$). Up to an irrelevant global phase, this condition can be achieved using $\mathcal{A} = \exp{-i \theta \mathcal{R}_{y}}$, where $\theta=\pi$ and $\mathcal{R}_{y} = (\nicefrac{1}{2})\sum_{j=1}^{n} \mathbb{1}^{\otimes j-1} \otimes \sigma^{y}_{j} \otimes \mathbb{1}^{\otimes n-j}$. The operation $\mathcal{M}(\phi)=\left[\exp{-i\phi\mathscr{S}}\right]^{r}$ is now a collective $z$-rotation generated by the operator $\mathscr{S}=(\nicefrac{1}{2})\sum_{j=1}^{n}\mathbb{1}^{\otimes j-1} \otimes \sigma^{z}_{j} \otimes \mathbb{1}^{\otimes n-j}$. Under such conditions, Eq.~\eqref{eq:QFI pure state protocol supp} becomes 
\begin{equation}
    \label{eq:QFI collective supp}
    \mathcal{I}(\phi)=4r^{2}\left(\langle\mathscr{S}^{2}\rangle-\langle\mathscr{S}\rangle^{2}\right),\qquad \langle\,\cdot\,\rangle\coloneqq\bra{\chi}\,\cdot\,\ket{\chi},
\end{equation}
where here $\ket{\chi}=\mathcal{A}_{\mathcal{U}}\ket{0^{\otimes n}}$. Differently from the previous derivation here $\mathscr{S}\ket{0^{\otimes n}}=(\nicefrac{n}{2})\ket{0^{\otimes n}}$, $\Tr\mathscr{S}=0$ and $\Tr\mathscr{S}^{2}=\nicefrac{n 2^n}{4}$, but a similar decomposition as Eq.~\eqref{eq:decomp supp} still holds, allowing us to write
\begin{equation}
    \label{eq:eq:decomp supp global}
    \langle\mathscr{S}\rangle=X^{2}s+(1-X^{2})\,Y,\qquad
    \langle\mathscr{S}^{2}\rangle=X^{2}s^{2}+(1-X^{2})\,W \, ,
\end{equation}
where again the subspace $\mathscr{H}_{0^{\otimes n}}^{\perp}$ is invariant under $\mathscr{S}$ and now $Y\coloneqq\bra{\mu}\Tilde{\mathscr{S}}\ket{\mu}$. The main differences compared to the previous decomposition are the terms $W\coloneqq\bra{\mu}\Tilde{\mathscr{S}}^{2}\ket{\mu}$ and the presence of the eigenvalue $s=\nicefrac{n}{2}$. Because of the new definition of the operator $\mathscr{S}$, we also have new trace properties, as $\Tr\Tilde{\mathscr{S}}=-s$ and $\Tr\Tilde{\mathscr{S}}^{2}=d\left( \frac{n}{4} \right)-\left( \frac{n}{2} \right)^{2}$. With these new properties, according to Eq.~\eqref{eq:Haar random state E averages supp}, with $d'=d-1$, we obtain
\begin{equation}
    \label{eq:EY collective supp}
    \underset{{\zeta \sim H}}{\mathbb{E}}[Y]=-\frac{\left( \frac{n}{2} \right)}{d-1},\qquad
    \underset{{\zeta \sim H}}{\mathbb{E}}[Y^{2}]=\frac{\left(\frac{n}{4}\right)}{d-1},\qquad
    \underset{{\zeta \sim H}}{\mathbb{E}}[W]=\frac{d \left(\frac{n}{4}\right)-\left( \frac{n}{2} \right)^{2}}{d-1}.
\end{equation}
According to the decomposition presented in Eq.~\eqref{eq:decomp supp} and the corresponding results presented in Eq.~\eqref{eq:aux1 supp},
\begin{align}
    \label{eq:Haar collective moments supp}
    \underset{{\mathcal{U} \sim H}}{\mathbb{E}}\left[\langle\mathscr{S}\rangle\right]&=-\frac{s}{(d-1)(d+1)},\notag\\
    \underset{{\mathcal{U} \sim H}}{\mathbb{E}}\left[\langle\mathscr{S}^{2}\rangle\right]&=\frac{d^{2}\left( \frac{n}{4} \right)-s^{2}}{(d-1)(d+1)},\\
    \underset{{\mathcal{U} \sim H}}{\mathbb{E}}\left[\langle\mathscr{S}\rangle^{2}\right]&=\frac{s^{2}(d-3)+\left( \frac{n}{4} \right)\,d(d+2)}{(d-1)(d+1)(d+3)},\notag
\end{align}
where we have again used the results obtained in the previous section, namely $\mathbb{E}[X^{2}]=\tfrac{1}{d+1}$, $\mathbb{E}[X^{4}]=\tfrac{3}{(d+1)(d+3)}$. Now it is possible to calculate the ensemble average of the QFI as
\begin{equation}
    \label{eq:QFI collective haar supp}
    \underset{{\mathcal{U} \sim H}}{\mathbb{E}}\left[\mathcal{I}(\phi)\right]= 4r^{2} \left( \underset{{\mathcal{U} \sim H}}{\mathbb{E}}\left[\langle\mathscr{S}^{2}\rangle\right] - \underset{{\mathcal{U} \sim H}}{\mathbb{E}}\left[\langle\mathscr{S}\rangle^{2}\right] \right) =r^{2}\,\frac{n\,d\,(d^{2}+2d-2-2n)}{(d^{2}-1)(d+3)} \underset{\text{large } n}{\approx} n\,r^{2}.
\end{equation}
As a comparison, we report the results for the Clifford random circuit case again. Here, we provide a demonstration based on the eigenstates of the generator $\mathscr{S}$. For Clifford-random circuits, up to a global phase, $\mathcal{A}_{\mathcal{U}}=\pm P$ with $P$ a non-identity Pauli string, so $\ket{\chi}$ is, up to a global phase, a computational-basis state and therefore an eigenstate of $\mathscr{S}$. In particular, we have that 
\begin{equation}
    \label{eq:eigenstates S}
    \mathscr{S} \ket{\chi} = \mathscr{S} \mathcal{A}_{\mathcal{U}}\ket{0^{\otimes n}} = \mathscr{S} \mathcal{U}^{\dagger} \mathcal{A} \mathcal{U} \ket{0^{\otimes n}} = \left( \frac{n}{2} - N_{\ket{1}} \right) \ket{\chi} \, ,
\end{equation}
where $N_{\ket{1}}$ represents the number of qubits in state $\ket{1}$ after the application of $\mathcal{A}_{\mathcal{U}}$ to the initial state $\ket{0^{\otimes n}}$. Independently of the value of $\left( \frac{n}{2} - N_{\ket{1}} \right)$ the state $\ket{\chi}$ is an eigenstate of $\mathscr{S}$, and consequently, according to Eq.~\eqref{eq:QFI collective supp} the QFI is identically null for every number of qubits $n$,
\begin{equation}
    \label{supp:eq:QFI collective clifford supp}
    \underset{{\mathcal{U} \sim C}}{\mathbb{E}}\left[\mathcal{I}(\phi)\right]=0 \, .
\end{equation}
Similar results may be found using the reasoning done in the previous section and considering the commuting and anti-commuting properties of Pauli strings (see Eq.~\eqref{eq:OTOC clifford ensemble average supp} and subsequent discussion). Again, rather than the exact formula, we are interested in the scaling and in the difference in the QFI for large $n$. Specifically, the scaling of the QFI for large $n$ is $\approx nr^2$ for Haar and $=0$ for Clifford. This means that it is possible to distinguish between the two cases, and the distinction becomes more and more pronounced as the number of qubits considered increases. 

\subsubsection*{An Alternative Global-Control Protocol: Parity Encoding}

After defining protocols in which qubits may be assessed individually and in which qubits are globally controlled, we propose a second scheme in which only global operations on all qubits are allowed. In particular, we employ a first global operation $\mathcal{A} = \exp{-i \theta \mathcal{R}_{y}}$, where $\theta=\pi$ and $\mathcal{R}_{y} = (\nicefrac{1}{2})\sum_{j=1}^{n} \mathbb{1}^{\otimes j-1} \otimes \sigma^{y}_{j} \otimes \mathbb{1}^{\otimes n-j}$ and here $\mathcal{M}(\phi) = p^{r}(\phi) = \left[\exp{-i \phi \mathscr{P}} \right]^{r}$, where $\mathscr{P} = \otimes_{j=1}^{n} \sigma^{z}_{j}$. Related parity-based operations have been studied in quantum computing~\cite{riste2013deterministic,fellner2022applications,fellner2022universal}, while a scheme for implementing parity phase gates through global phase modulation of a Rydberg excitation laser has been proposed for neutral-atom systems~\cite{kazemi2025multiqubit}. These operations lead to a QFI in the form:
\begin{equation}
    \label{eq:OTOC QFI Zn supp}
    \mathcal{I}(\phi) = 4 r^2 \left( 1 - \langle \mathscr{P}^{\dagger} \mathscr{A}_{\mathcal{U}}^{\dagger} \mathscr{P} \mathscr{A}_{\mathcal{U}} \rangle^2 \right) \, ,
\end{equation}
where $\mathscr{A}_{\mathcal{U}} = \mathcal{U}^{\dagger} \mathcal{A} \mathcal{U}$. When unitaries $\mathcal{U}$ and $\mathcal{U}^{\dagger}$ are generated by sufficiently deep random circuits that resemble the Haar statistics, the ensemble averages reproduce those obtained with local control, and consequently the ensemble-averaged QFI is given by Eq.~\eqref{eq:OTOC QFI Zn haar supp}, up to a multiplicative factor of 4. The results are valid for every non-trivial Pauli string, whether it acts on one qubit ($\mathscr{S}_{j}$) or all $n$ ($\mathscr{P}$), since the demonstration relies only on the properties $\mathscr{S}_{j}^{2} = \mathbb{1}$ and $\Tr{\mathscr{S}_{j}} = 0$, which are valid also for $\mathscr{P}$, i.e., $\mathscr{P}^{2} = \mathbb{1}$ and $\Tr{\mathscr{P}} = 0$. Indeed, independently of the support of the operation (single qubit for $\mathscr{S}_{j}$ or all qubits for $\mathscr{P}$), the ensemble average results of deep quantum random circuits are the same. As for the Haar-random case, also in the limit of Clifford-random circuits, the results resemble those presented in ``Characterization through Local Control", as the results do not depend on the support of the operators. Following the same considerations made in the previous subsection, $\langle \mathscr{P}^{\dagger} \mathscr{A}_{\mathcal{U}}^{\dagger} \mathscr{P} \mathscr{A}_{\mathcal{U}} \rangle = \pm 1$, because operators $\mathscr{P}$ and $\mathscr{A}_{\mathcal{U}}$ commute with probability $p_{c} = \nicefrac{(2^{2n -1}-1)}{4^{n} -1}$ or anti-commute with probability $p_a = \nicefrac{(2^{2n -1})}{4^{n} -1}$. Then $\underset{{\mathcal{U} \sim C}}{\mathbb{E}} \left[ \langle \mathscr{P}^{\dagger} \mathscr{A}_{\mathcal{U}}^{\dagger} \mathscr{P} \mathscr{A}_{\mathcal{U}} \rangle \right] = -\frac{1}{d^{2}-1}$, and $\underset{{\mathcal{U} \sim C}}{\mathbb{E}} \left[ \langle \mathscr{P}^{\dagger} \mathscr{A}_{\mathcal{U}}^{\dagger} \mathscr{P} \mathscr{A}_{\mathcal{U}} \rangle^2 \right] = +1$. Consequently, the ensemble average of Eq.~\eqref{eq:OTOC QFI Zn supp} is the same as Eq.~\eqref{eq:OTOC QFI Zn clifford supp}, i.e., $\underset{{\mathcal{U} \sim C}}{\mathbb{E}} \left[ \mathcal{I}(\phi) \right] = 0$. The intermediate scenario, in which Eq.~\eqref{eq:OTOC QFI Zn supp} is neither minimized nor maximized, corresponds to intermediate scenarios in which the information scrambling partially reproduces Haar-random ensemble averages. These results demonstrate that we have found an efficient protocol for benchmarking random quantum circuits on platforms that support the required preparation and perturbation operations.

\paragraph{Measurement} After the definition of the two protocols, here we propose a few suitable techniques for implementing the final measurement procedure. The first possible choice would be the projective measurement in the computational basis after the application of a unitary operation $\left(\mathcal{M}(\phi_{0})\mathcal{A}_{\mathcal{U}}\right)^{\dagger}$, with $\phi_{0} \approx \phi$ \footnote{When there is a repetition of the application of an operator, as for example in the case of the exact local protocol, the approximation reads $r \phi_{0} \approx r \phi$.}. Defining $\ket{\psi_0'}=\ket{\psi(\phi_0)}$ we will consider the binary measurement $\{\Pi_0,\mathbb{1}-\Pi_0\}$ with $\Pi_0=\ketbra{\psi_0'}{\psi_0'}$. Defining $\delta=\phi-\phi_0$ with $\phi_0$ held fixed, we can write $\ket{\psi(\phi)}=e^{-i\delta\mathcal{S}}\ket{\psi_0'}$, where $\mathcal{S}$ denotes the full encoding generator, including the repetition factor $r$. The probability of measuring the initial state $\ket{\psi_{0}} = \ket{0^{\otimes n}}$ is
\begin{equation}
  p(\phi)=\left|\bra{\psi_0'}e^{-i\delta \mathcal{S}}\ket{\psi_0'}\right|^2 = \left|\langle e^{-i\delta \mathcal{S}}\rangle\right|^2,
\end{equation}
where $\langle\cdot\rangle$ denotes the expectation value on $\ket{\psi_0'}$. Expanding the amplitude for small $\delta$, $\langle e^{-i\delta \mathcal{S}}\rangle = 1-i\delta\langle \mathcal{S}\rangle-\frac{\delta^2}{2}\langle \mathcal{S}^2\rangle+O(\delta^3)$ we can rewrite the probability $p(\phi)$ as
\begin{equation}
\label{eq:probability supp}
  p(\phi)=1-\delta^2\,\operatorname{\operatorname{Var}}(\mathcal{S})+O(\delta^3).
\end{equation}
It follows that $\partial_\phi p=-2\delta\,\operatorname{\operatorname{Var}}(\mathcal{S})+O(\delta^2)$ and $p(1-p)=\delta^2\,\operatorname{\operatorname{Var}}(\mathcal{S})+O(\delta^3)$. The classical Fisher information of this binary measurement is therefore
\begin{equation}
\label{eq:variance supp 2}
  F_C(\phi)=\frac{(\partial_\phi p(\phi))^2}{p(\phi)(1-p(\phi))}\;\xrightarrow{\;\delta\to0\;}\;4\,\operatorname{\operatorname{Var}}(\mathcal{S})=\mathcal{I}(\phi).
\end{equation}
When $\operatorname{\operatorname{Var}}(\mathcal{S})=0$, the state changes only by an overall phase, and both the classical FI and QFI vanish. Alternatively, since Eqs.~\eqref{eq:OTOC QFI single pauli supp} and~\eqref{eq:OTOC QFI Zn supp} depend only on $\langle \mathscr{P}^{\dagger} \mathscr{A}_{\mathcal{U}}^{\dagger} \mathscr{P} \mathscr{A}_{\mathcal{U}} \rangle^2$ or $\langle \mathscr{S}_{j}^{\dagger} \mathscr{A}_{k,\mathcal{U}}^{\dagger} \mathscr{S}_{j} \mathscr{A}_{k,\mathcal{U}} \rangle^2$, a suitable measurement choice to infer information about the QFI in the local and parity protocols is to study the OTOC on which the QFI depends, using the standard protocols adopted in the measurement of information scrambling, from projective measurement over the initial state to interferometric procedures (see Sec.~\ref{sec:A Brief Overview of Out-of-Time Order Correlations}).

\subsection{Benchmarking with Single-Qubit Readout}

To conclude the explicit derivation of the protocols presented in the main text, we also present the detailed calculations related to the perturbative ``butterfly" protocol. This procedure is complementary with respect to the two previous ones, as it requires only a final local measurement on a single qubit. Mathematically, we analyze the inverse sensitivity and its variance of a state in the form $\ket{\psi(\phi)} = \left[ \mathcal{U} \mathcal{M}(\phi) \mathcal{U}^{\dagger} \left( \frac{\mathbb{1} + i\mathscr{V}}{\sqrt{2}} \right) \mathcal{U} \right] \ket{0^{\otimes n}}$, and conclude the protocol with a local measurement on $\mathscr{V}$. Here $\mathcal{M}(\phi) = \exp{(-i\phi \mathcal{S})}$, with $\mathcal{S} = (\nicefrac{1}{2})\sum_{j=1}^{n} \mathbb{1}^{\otimes j-1} \otimes \sigma^{z}_{j} \otimes \mathbb{1}^{\otimes n-j}$. A similar procedure, focused on creating a suitable metrological state for sensing an external magnetic field, was first proposed theoretically in \cite{kobrin2024universal} and subsequently experimentally verified in \cite{hu2026quantum}, thereby demonstrating the experimental feasibility of the present protocol. The focus of our protocol is distinct and goes beyond the theoretical calculation performed, as, through the inverse sensitivity and its variance of the created state, we can characterize the randomness of the unitary $\mathcal{U}$. Being the final measurement, a local measurement of the operator $\mathscr{V}$, we are interested in the expectation value $\langle \mathscr{V} \rangle_{\phi}$, as the classical Fisher information of the measurement is $F^{\psi}_{\mathscr{V}}(\phi) = \nicefrac{\left( \partial_{\phi} \langle \mathscr{V} \rangle_{\phi} \right)^2 }{\left( 1 - \langle \mathscr{V} \rangle_{\phi}^2 \right)}$. We are then interested in the expectation value of $\langle \mathscr{V} \rangle_{\phi}$. Following the calculations presented in \cite{kobrin2024universal}, in the weak signal regime, i.e., when $\phi \approx 0$, the expectation value $\langle \mathscr{V} \rangle_{\phi}$ may be written as $\langle \mathscr{V} \rangle_{\phi} \approx \phi \left( \frac{n}{2} - \bra{0^{\otimes n}} \mathscr{V}_{\mathcal{U}} \mathcal{S} \mathscr{V}_{\mathcal{U}} \ket{0^{\otimes n}}  \right)$. In such conditions, the inverse sensitivity of the system can be written as \footnote{This is an approximation that holds when $F^{\psi}_{\mathscr{V}}(\phi) = \nicefrac{\left( \partial_{\phi} \langle \mathscr{V} \rangle_{\phi} \right)^2 }{\left( 1 - \langle \mathscr{V} \rangle_{\phi}^2 \right)} \approx \left( \partial_{\phi} \langle \mathscr{V} \rangle_{\phi} \right)^2$. When this approximation does not hold, the following results are related only to the response slope magnitude $\left| \partial_{\phi} \langle \mathscr{V} \rangle_{\phi} \right|$, and not to the overall sensitivity $\eta^{-1}_{\phi \approx 0}$.}
\begin{align}
    \label{eq:sensitivity weak signal regime supp}
    \eta^{-1}_{\phi \approx 0} & \approx \left( \frac{n}{2} - \bra{0^{\otimes n}} \mathscr{V}_{\mathcal{U}} \mathcal{S} \mathscr{V}_{\mathcal{U}} \ket{0^{\otimes n}}  \right) = \frac{1}{2} \sum_{i=1}^{n} \left( 1 - \bra{0^{\otimes n}} \sigma_{i}^{z} \mathscr{V}_{\mathcal{U}} \sigma_{i}^{z} \mathscr{V}_{\mathcal{U}} \ket{0^{\otimes n}} \right) \, .
\end{align}
Since this equation depends on the sum of OTOCs in the form $\bra{0^{\otimes n}} \sigma_{i}^{z} \mathscr{V}_{\mathcal{U}} \sigma_{i}^{z} \mathscr{V}_{\mathcal{U}} \ket{0^{\otimes n}}$, where there are only 2 copies of $\mathcal{U}$ and $\mathcal{U}^{\dagger}$, Clifford-random and Haar-random statistics lead to the same ensemble average of the inverse sensitivity. Explicitly, according to the calculations made before, for each OTOC $\underset{{\mathcal{U} \sim H}}{\mathbb{E}} \left[ \bra{0^{\otimes n}} \sigma_{i}^{z} \mathscr{V}_{\mathcal{U}} \sigma_{i}^{z} \mathscr{V}_{\mathcal{U}} \ket{0^{\otimes n}} \right]=\underset{{\mathcal{U} \sim C}}{\mathbb{E}} \left[ \bra{0^{\otimes n}} \sigma_{i}^{z} \mathscr{V}_{\mathcal{U}} \sigma_{i}^{z} \mathscr{V}_{\mathcal{U}} \ket{0^{\otimes n}} \right]= -\frac{1}{d^2 - 1}$, and then
\begin{equation}
    \label{eq:ensamble average eta supp}
    \underset{{\mathcal{U} \sim H}}{\mathbb{E}} \left[ \eta^{-1}_{\phi \approx 0} \right] = \underset{{\mathcal{U} \sim C}}{\mathbb{E}} \left[ \eta^{-1}_{\phi \approx 0} \right] = \frac{\log_{2}{d}}{2} \left( 1 + \frac{1}{d^2-1} \right)  \, ,
\end{equation}
where the prefactor $\log_{2}{d}$ arises from the sum of the $n=\log_{2}{d}$ different OTOCs. To recover the difference between the statistics of Clifford-random and Haar-random circuits, we have to go beyond the mean inverse sensitivity and examine its variance, as it involves OTOCs with $4$ copies of $\mathcal{U}$ and $\mathcal{U}^{\dagger}$. To calculate the variance of Eq.~\eqref{eq:sensitivity weak signal regime supp}, we have to rely on the covariance matrix of the different OTOCs
\begin{equation}
    \label{eq:covariance matrix supp}
    \operatorname{\operatorname{Var}} (\eta^{-1}_{\phi \approx 0}) = \frac{1}{4} \left( \sum_{i,j}^{n} \operatorname{Cov}( O_{i} , O_{j} ) \right) \, ,
\end{equation}
where $O_{i} = \bra{0^{\otimes n}} \sigma_{i}^{z} \mathscr{V}_{\mathcal{U}} \sigma_{i}^{z} \mathscr{V}_{\mathcal{U}} \ket{0^{\otimes n}} $ and $O_{j} = \bra{0^{\otimes n}} \sigma_{j}^{z} \mathscr{V}_{\mathcal{U}} \sigma_{j}^{z} \mathscr{V}_{\mathcal{U}} \ket{0^{\otimes n}}$. Following the symmetric decomposition presented in Eq.~\eqref{eq:decomp supp}, $\ket{\chi}=X\ket{0^{\otimes n}}+\sqrt{1-X^{2}}\,\ket{\mu},\quad \ket{\mu}\in\mathscr{H}_{0^{\otimes n}}^\perp$, we can explicitly calculate each element of $\operatorname{Cov}( O_{i} , O_{j} )$ in the Haar-random limit. Specifically, according to the properties $\Tr{\Tilde{\sigma_{i}^{z}}} = \Tr{\Tilde{\sigma_{j}^{z}}} = -1$ and $\Tr{\Tilde{\sigma_{i}^{z}}\Tilde{\sigma_{j}^{z}}}= -1$ for $i\neq j$, where $\Tilde{\sigma_{j}^{z}}$ acts only on $\mathscr{H}_{0^{\otimes n}}^\perp$, and because $\mathbb{E}[Y_{i} Y_{j}] = 0$ for $i\neq j$, with 
$Y_{i}\coloneqq\bra{\mu}\Tilde{\sigma_{i}^{z}}\ket{\mu}$, each useful ensemble average reads
\begin{align}
    \label{eq:cov haar supp}
    \underset{{\mathcal{U} \sim H}}{\mathbb{E}} \left[ O_{i} \right] &= -\frac{1}{d^2 -1} \, , \\
    \underset{{\mathcal{U} \sim H}}{\mathbb{E}} \left[ O_{i}^2 \right] &= \frac{d^2 + 3d -3}{(d-1)(d+1)(d+3)} \, , \\
    \underset{{\mathcal{U} \sim H}}{\mathbb{E}} \left[ O_{i} O_{j} \right] &= \frac{d - 3}{(d-1)(d+1)(d+3)}, \qquad i\neq j \, .
\end{align}
In the limit of the Haar-random ensemble average, the variance of the inverse sensitivity is then
\begin{align}
    \label{eq:variance sensitivity Haar supp}
    \underset{{\mathcal{U} \sim H}}{\operatorname{Var}} \left( \eta^{-1}_{\phi \approx 0} \right) = \frac{\log_{2}{d}}{4} \bigg[& \frac{d^2 + 3d - 3 + (\log_{2}(d)-1)(d-3)}{(d-1)(d+1)(d+3)} - \frac{\log_{2}(d)}{(d^2-1)^2} \bigg] \, .
\end{align}
In contrast, in the limit of Clifford-random circuits, each useful ensemble average reads
\begin{align}
    \label{eq:cov clifford supp}
    \underset{{\mathcal{U} \sim C}}{\mathbb{E}} \left[ O_{i} \right] &= -\frac{1}{d^2 -1} \, , \\
    \underset{{\mathcal{U} \sim C}}{\mathbb{E}} \left[ O_{i}^2 \right] &= 1 \, , \\
    \underset{{\mathcal{U} \sim C}}{\mathbb{E}} \left[ O_{i} O_{j} \right] &= -\frac{1}{d^2 -1}, \qquad i\neq j \, .
\end{align}
Then, $\underset{{\mathcal{U} \sim C}}{\operatorname{Var}} \left( \eta^{-1}_{\phi \approx 0} \right)$ is 
\begin{equation}
    \label{eq:variance sensitivity Clifford supp}
    \underset{{\mathcal{U} \sim C}}{\operatorname{Var}} \left( \eta^{-1}_{\phi \approx 0} \right) = \frac{\log_{2}{d}}{4} \left[  \frac{d^2 (d^2 - \log_{2}(d) - 1)}{(d^2 - 1)^2} \right] \, .
\end{equation}
This means that the variance of the inverse sensitivity scales approximately linearly with the number of qubits and can be measured on quantum platforms with many qubits. This result is in contrast to Eq.~\eqref{eq:variance sensitivity Haar supp}, where the variance decreases with the number of qubits, approaching the limit $\underset{{\mathcal{U} \sim H}}{\operatorname{Var}} \sim \nicefrac{\log_{2}{d}}{4d} = \nicefrac{n}{4\cdot 2^n} \approx 0$ for large quantum platforms. Also, the present protocol is suitable for characterizing randomness in current quantum platforms and, in particular, will become increasingly favorable as the number of qubits implemented in quantum processors increases. As for local and global exact protocols, also in the butterfly benchmarking protocol, the difference between Haar- and Clifford-random is particularly evident. This immediately leads to the same considerations expressed earlier, with the possibility of identifying intermediate scenarios in which the circuits partially resemble Haar-random statistics, with a scaling of the variance between $\sim \nicefrac{\log_{2}{d}}{4d}$ and $\sim \nicefrac{\log_{2}{d}}{4}$. 

\section{White Noise Quantum Metrology}
\label{sec:White Noise Quantum Metrology}

In this section, we present theoretical results for any metrological protocol subject to white noise. This corresponds to a standard benchmark for noise resilience in quantum circuits, since random quantum circuits, under specific conditions, transform local noise into global white noise \cite{arute2019quantum,dalzell2024random}. As reported in the main text, this analysis serves as a benchmark noise model rather than a universal description of experimental noise and is used to test the robustness of our protocol to experimental imperfections. First, we introduce the general form of the density matrix, as
\begin{equation}
    \label{eq:noisy rho supp}
    \varrho = \mathscr{F} \ket{\psi(\phi)} \bra{\psi(\phi)} + (1-\mathscr{F}) \varrho_{w.n.} \, ,
\end{equation}
where $\ket{\psi(\phi)}$ is the pure state that encodes information about the parameter $\phi$ subject to the estimation protocol and $\varrho_{w.n.} = \nicefrac{\mathbb{1}}{2^n}$ is the maximally mixed state in a Hilbert space of dimension $d=2^n$, and does not depend on the parameter $\phi$. $\mathscr{F}$ is the $\phi$-independent weight of the ideal pure-state component, with $0 \leq \mathscr{F} \leq 1$, and is related to the purity of the state $\gamma$ through the relation $\gamma = \Tr{\varrho^2} = \mathscr{F}^2 + 2^{-n}(1-\mathscr{F}^{2})$. Under these conditions, the derivative of the density matrix $\varrho$ is as follows:
\begin{align}
    \label{eq:derivative of rho white noise supp}
    \partial \varrho = \partial \left( \mathscr{F} \ket{\psi(\phi)} \bra{\psi(\phi)} + (1-\mathscr{F}) \varrho_{w.n.} \right) = \mathscr{F} \partial_{\phi} \left(\ket{\psi(\phi)} \bra{\psi(\phi)}\right) \, .
\end{align}
According to the definition of QFI in terms of SLDs $\partial \varrho = \frac{1}{2} \{ \mathcal{L}, \varrho \}$, we can write
\begin{align}
    \label{eq:QFI SLDs white noise supp}
    \mathscr{F} \partial_{\phi} \left(\ket{\psi(\phi)} \bra{\psi(\phi)}\right) = \frac{1}{2} \{ \mathcal{L}^{\varrho} , \mathscr{F} \ket{\psi(\phi)} \bra{\psi(\phi)}  \} + \frac{1}{2} \{ \mathcal{L}^{\varrho} , (1-\mathscr{F}) \varrho_{w.n.} \} \, .
\end{align}
Since the LHS of Eq.~\eqref{eq:QFI SLDs white noise supp} is proportional to $\{\mathcal{L}^{\psi}, \ket{\psi(\phi)} \bra{\psi(\phi)} \}$ and $\varrho_{w.n.} = \nicefrac{\mathbb{1}}{2^{n}}$, it is reasonable to consider $\mathcal{L}^{\varrho} = \alpha \mathcal{L}^{\psi}$. Substituting these considerations into Eq.~\eqref{eq:QFI SLDs white noise supp}, and considering $\{ \mathcal{L}^{\psi} , \varrho_{w.n.} \} = 2 \mathcal{L}^{\psi}  \varrho_{w.n.} = 2 \varrho_{w.n.} \mathcal{L}^{\psi} $, we obtain the following relation
\begin{align}
    \label{eq:QFI SLDs white noise ansatz supp}
    \frac{\mathscr{F}}{2} \{ \mathcal{L}^{\psi} , \ket{\psi(\phi)} \bra{\psi(\phi)} \} = \alpha\frac{ \mathscr{F}}{2} \{ \mathcal{L}^{\psi} , \ket{\psi(\phi)} \bra{\psi(\phi)} \} + \alpha\frac{\left( 1 - \mathscr{F} \right)}{2^{n}} \mathcal{L}^{\psi} \, .
\end{align}
Observing that $\{ \mathcal{L}^{\psi}, \ket{\psi(\phi)} \bra{\psi(\phi)} \} = \mathcal{L}^{\psi}$ this directly gives us a condition on the constant factor $\alpha$ and consequently the form of the SLD for the noisy case, $\mathcal{L}^{\varrho}$, in terms of the pure state case, $\mathcal{L}^{\psi}$, as
\begin{equation}
    \label{eq:optimal noisy measurement supp}
    \mathcal{L}^{\varrho} = \left( \frac{\mathscr{F}}{\mathscr{F} + \frac{(1-\mathscr{F})}{2^{n-1}}} \right) \mathcal{L}^{\psi} \, ,
\end{equation}
i.e., SLD eigenbasis is unchanged; only its eigenvalues are rescaled. Substituting Eq.~\eqref{eq:optimal noisy measurement supp} into the relationship between SLD and QFI, $\mathcal{I}^{\varrho} (\phi) = \Tr{ \varrho (\mathcal{L}^{\varrho})^2 }$, we can obtain the QFI as
\begin{align}
    \label{eq:noisy qfi supp}
    \mathcal{I}^{\varrho} (\phi) & = \left( \frac{\mathscr{F}}{\mathscr{F} + \frac{(1-\mathscr{F})}{2^{n-1}}} \right)^2 \Tr{ \left( \mathscr{F} \ket{\psi(\phi)} \bra{\psi(\phi)} + (1-\mathscr{F}) \varrho_{w.n.} \right) \left[ 2 \left( \ket{\psi(\phi)} \bra{\partial \psi(\phi)} + \ket{\partial \psi(\phi)} \bra{\psi(\phi)} \right) \right]^2 } =  \nonumber \\
    &= \left( \frac{\mathscr{F}}{\mathscr{F} + \frac{(1-\mathscr{F})}{2^{n-1}}} \right)^2 \left[ \mathscr{F} \mathcal{I}^{\psi} (\phi) + 4 \frac{(1-\mathscr{F})}{2^n} \Tr{ \left( \ket{\psi(\phi)} \bra{\partial \psi(\phi)} + \ket{\partial \psi(\phi)} \bra{\psi(\phi)} \right)^2} \right] = \nonumber \\
    &= \left( \frac{\mathscr{F}}{\mathscr{F} + \frac{(1-\mathscr{F})}{2^{n-1}}} \right)^2 \left[ \mathscr{F} \mathcal{I}^{\psi} (\phi) + 8 \frac{(1-\mathscr{F})}{2^n} \left( \abs{\partial \psi (\phi)}^2 + \bra{\psi(\phi)} \ket{\partial \psi (\phi)}^2 \right)  \right] = \nonumber  \\
    &= \left( \frac{\mathscr{F}}{\mathscr{F} + \frac{(1-\mathscr{F})}{2^{n-1}}} \right)^2 \left[ \mathscr{F} \mathcal{I}^{\psi} (\phi) + 2 \frac{(1-\mathscr{F})}{2^n} \mathcal{I}^{\psi}   \right] = \left( \frac{\mathscr{F}^2}{\mathscr{F} + \frac{(1-\mathscr{F})}{2^{n-1}}} \right) \mathcal{I}^{\psi} (\phi) \, .
\end{align}
Indeed, the QFI $\mathcal{I}^{\varrho} (\phi)$ is proportional to $\mathcal{I}^{\psi} (\phi)$. After obtaining the exact results for the optimal measurement and the exact QFI for every quantum metrology protocol in the presence of white noise, we analyze the limit $d\mathscr{F} \gg 2(1-\mathscr{F})$. The SLD can be approximated as $\mathcal{L}^{\varrho} \simeq \mathcal{L}^{\psi}$, i.e., the same as the non-noisy case. Additionally, the QFI follows the relation $\mathcal{I}^{\varrho} (\phi) \simeq \mathscr{F} \mathcal{I}^{\psi} (\phi)$, and is then simply the one obtained in the non-noisy case multiplied by the weight $\mathscr{F}$. Importantly, this means that in the noisy case, all the results obtained in terms of the QFI are simply proportional to the ones obtained in Sec.~\ref{sec:Calculation Details}. Then, if the QFI is exactly zero, as in the Clifford-random case of the non-noisy exact protocol, it remains zero in the presence of white noise. The size dependence of the noisy QFI is determined jointly by the ideal QFI and the noise-dependent prefactor, including any dependence of $\mathscr{F}$ on the system size.

We stress that the proportionality relations obtained in the limit of large Hilbert spaces for Eqs.~\eqref{eq:optimal noisy measurement supp}-\eqref{eq:noisy qfi supp} are valid only for the SLD and the QFI, and are not generally true for any measurement procedure or for any FI. Indeed, for a POVM $\{\Pi_{x}\}$ the conditional probability $p^{\varrho}\left( x \vert \phi \right) = \Tr{\varrho \Pi_{x} }$, where $p^{\varrho}\left( x \vert \phi \right) = \mathscr{F} p^{\psi}\left( x \vert \phi \right) + 2^{-n} (1- \mathscr{F}) \Tr{\Pi_{x}}$ and consequently $\partial p^{\varrho}\left( x \vert \phi \right) = \mathscr{F} \partial p^{\psi}\left( x \vert \phi \right)$. The FI is then 
\begin{equation}
\label{eq:white noise fi supp}
    F^{\varrho}_{x}(\phi) = \sum_{x} \dfrac{\left| \partial_{\phi} p^{\varrho} \left( x \vert \phi \right) \right|^{2}}{p^{\varrho} \left( x \vert \phi \right)} = \sum_{x} \dfrac{ \mathscr{F}^{2} \left| \partial_{\phi} p^{\psi} \left( x \vert \phi \right) \right|^{2}}{\mathscr{F} p^{\psi} \left( x \vert \phi \right) + 2^{-n} (1-\mathscr{F}) \Tr{\Pi_{x}} } \, .
\end{equation}
The FI of the noisy density matrix $\varrho$ is proportional to the case of pure state when the POVM is the projective measurement in the eigenbasis of the SLD defined in Eq.~\eqref{eq:optimal noisy measurement supp}, as $F^{\varrho, opt}_{x}(\phi) = \mathcal{I}^{\varrho} (\phi)$. In the limit $\mathscr{F} p^{\psi} \left( x \vert \phi \right) \gg 2^{-n} (1-\mathscr{F}) \Tr{\Pi_{x}} $, $F^{\varrho}_{x}(\phi) \simeq \mathscr{F} F^{\psi}_{x}(\phi)$, similar to the effective proportionality of the QFI. These results apply to arbitrary pure probe states and physical platforms within the assumed global white-noise model. This perfectly matches the scope of our paper, which proposes a protocol with global or local operations and measurements that account for noise effects. Comparing experimental results with Eqs.~\eqref{eq:optimal noisy measurement supp}, \eqref{eq:noisy qfi supp}, and \eqref{eq:white noise fi supp} provides a test of consistency with the assumed global white-noise model. In the benchmark with a single qubit readout according to the measurement $M=\mathscr{V}$, the FI of Eq.~\eqref{eq:FI supp} is reduced to
\begin{equation}
    \label{eq:single qubit fi}
    F^{\psi}_{\mathscr{V}}(\phi) = \frac{\left( \partial_{\phi} \langle \mathscr{V} \rangle_{\phi} \right)^2 }{\left( 1 - \langle \mathscr{V} \rangle_{\phi}^2 \right)} \, ,
\end{equation}
where the superscript $\psi$ refers to the ideal pure probe state. For white noise, this formula is reduced to
\begin{equation}
    \label{eq:single qubit fi white noise supp}
    F^{\varrho}_{\mathscr{V}}(\phi) = \frac{ \mathscr{F}^2 \left( \partial_{\phi} \langle \mathscr{V} \rangle_{\phi} \right)^2 }{\left( 1 - \mathscr{F}^2 \langle \mathscr{V} \rangle_{\phi}^2 \right)} \, .
\end{equation}
Then, Fisher information generally does not obey a simple scaling $\mathscr{F}^2$, but in the limit of $\langle \mathscr{V} \rangle_{\phi} \approx 0$ $F^{\varrho}_{\mathscr{V}}(\phi) \approx \mathscr{F}^2 F^{\psi}_{\mathscr{V}}(\phi)$. For a common $\mathscr{F}$ across circuit realizations, the ensemble mean and variance of the QFI are rescaled exactly by $\alpha\mathscr{F}$ and $(\alpha\mathscr{F})^2$, respectively. For the single-qubit readout in the regime $\langle\mathscr{V}\rangle_{\phi}\approx0$, the mean inverse sensitivity is approximately multiplied by $\mathscr{F}$ and its variance by $\mathscr{F}^2$.

\section{Tests for higher-order \texorpdfstring{$k$}{k}-designs and OTOCs}
\label{sec:Test for higher order k-designs and OTOCs}

To test higher-order $k$-designs or to evaluate higher-order OTOCs, we sequentially apply the protocol proposed in Sec.~\ref{sec:Calculation Details}. Mathematically, we perform on the initial state $\ket{0^{\otimes n}}$ the following operation:
\begin{equation}
    \label{eq:sequential protocol supp}
    \ket{\psi_{m}(\phi)} = \left[  \mathcal{M}(\phi) \mathcal{U}^{\dagger} \mathcal{A} \mathcal{U} \right]^{m} \ket{0^{\otimes n}} = \left[  \mathcal{M}(\phi) \mathcal{A}_{\mathcal{U}} \right]^{m} \ket{0^{\otimes n}} \, ,
\end{equation}
where we consider $\mathcal{A}^{2}= \mathbb{1}$, with $\mathcal{A} = \mathcal{A}^{\dagger}$ and $\mathcal{S} = \mathcal{S}^{\dagger}$. A similar approach, based on repeating the same operation multiple times, has been adopted in the study of Chaotic CFT dynamics \cite{shenker2014multiple,roberts2015localized} and in enhanced techniques for estimating unknown variables in quantum systems \cite{cavazzoni2025frequency}. In such a protocol, the QFI of the state $\ket{\psi_{m}(\phi)}$ can be written as 
\begin{equation}
    \label{eq:multiple butterfly}
    \mathcal{I}_{m}(\phi) = 4 \left( \bra{0^{\otimes n}} \left( \partial_{\phi} \left[  \mathcal{M}(\phi) \mathcal{A}_{\mathcal{U}} \right]^{m} \right)^{\dagger} \left( \partial_{\phi} \left[  \mathcal{M}(\phi) \mathcal{A}_{\mathcal{U}} \right]^{m} \right) \ket{0^{\otimes n}} - \abs{\bra{0^{\otimes n}} \left[  \mathcal{M}(\phi) \mathcal{A}_{\mathcal{U}} \right]^{m,\dagger} \left( \partial_{\phi} \left[  \mathcal{M}(\phi) \mathcal{A}_{\mathcal{U}} \right]^{m} \right) \ket{0^{\otimes n}}}^2 \right) \, .
\end{equation}
In general, the operators $\left[  \mathcal{M}(\phi) \mathcal{A}_{\mathcal{U}} \right]$ and $\partial_{\phi} \left[  \mathcal{M}(\phi) \mathcal{A}_{\mathcal{U}} \right]$ do not commute, and then the derivative $\partial_{\phi} \left[  \mathcal{M}(\phi) \mathcal{A}_{\mathcal{U}} \right]^{m}$ has the form $\partial_{\phi} \left[  \mathcal{M}(\phi) \mathcal{A}_{\mathcal{U}} \right]^{m} =\sum_{k=0}^{m-1} \left[  \mathcal{M}(\phi) \mathcal{A}_{\mathcal{U}} \right]^k (\partial_{\phi} \left[  \mathcal{M}(\phi) \mathcal{A}_{\mathcal{U}} \right]) \left[  \mathcal{M}(\phi) \mathcal{A}_{\mathcal{U}} \right]^{m-1-k}$. The case $m=1$ reproduces the results presented in the main text and in Sec.~\ref{sec:Calculation Details}. Taking into account the special case $m=2$, $\ket{\psi_2(\phi)} = \mathcal{M}(\phi) \mathcal{A}_{\mathcal{U}} \mathcal{M}(\phi) \mathcal{A}_{\mathcal{U}} \ket{0^{\otimes n}}$, and $\ket{\partial_{\phi} \psi_2(\phi)} = -i\mathcal{S} \mathcal{M}(\phi) \mathcal{A}_{\mathcal{U}} \mathcal{M}(\phi) \mathcal{A}_{\mathcal{U}} \ket{0^{\otimes n}} -i\mathcal{M}(\phi) \mathcal{A}_{\mathcal{U}} \mathcal{S} \mathcal{M}(\phi) \mathcal{A}_{\mathcal{U}} \ket{0^{\otimes n}}$. Since we are not interested in the estimation of $\phi$, but in the characterization of the unitaries that define $\ket{\psi_2(\phi)}$, we can focus on a specific value of $\phi$ and use it to characterize the QFI (or analogously the sensitivity) of the state. In particular, we further assume $\mathcal{S}^2=\mathbb{1}$ and choose $\phi=\nicefrac{\pi}{2}$, so that $\mathcal{M}\left(\frac{\pi}{2}\right) = -i \mathcal{S}$. At $\phi=\pi/2$, $\ket{\psi_2\left(\frac{\pi}{2}\right)} = -\mathcal{S}\mathcal{A}_{\mathcal{U}}\mathcal{S}\mathcal{A}_{\mathcal{U}}\ket{0^{\otimes n}}$, and $\ket{\partial_{\phi}\psi_2(\phi)}|_{\phi=\pi/2} = -i\mathcal{S} \left( \ket{\psi_2} - \ket{0^{\otimes n}} \right)$. Accordingly, the QFI reads 
\begin{align}
    \label{eq:QFI m=2 supp}
    &\mathcal{I}_{2}(\phi)\Big|_{\phi=\pi/2} = 4\Big[\, 2 + 2\,\mathrm{Re}\,\langle 0^{\otimes n}|\,\mathcal{A}_{\mathcal{U}}^{\dagger 2}\, \mathcal{S}\, \mathcal{A}_{\mathcal{U}}\, \mathcal{S}\, \mathcal{A}_{\mathcal{U}}\,|0^{\otimes n}\rangle - \Big( \langle 0^{\otimes n}|\,\mathcal{A}_{\mathcal{U}}^{\dagger}\, \mathcal{S}\, \mathcal{A}_{\mathcal{U}}\,|0^{\otimes n}\rangle + \langle 0^{\otimes n}|\,\mathcal{A}_{\mathcal{U}}^{\dagger}\, \mathcal{S}\, \mathcal{A}_{\mathcal{U}}^{\dagger}\, \mathcal{S}\, \mathcal{A}_{\mathcal{U}}\, \mathcal{S}\, \mathcal{A}_{\mathcal{U}}\,|0^{\otimes n}\rangle \Big)^{2} \Big] \nonumber \\
    &= 4\Big[\, 2 + 2\,\langle 0^{\otimes n}|\, \mathcal{S}^{\dagger}\, \mathcal{A}_{\mathcal{U}}^{\dagger} \, \mathcal{S}\, \mathcal{A}_{\mathcal{U}}\,|0^{\otimes n}\rangle - \frac{1}{s^2}\Big( \langle 0^{\otimes n}|\,\mathcal{S}^{\dagger}\,\mathcal{A}_{\mathcal{U}}^{\dagger}\, \mathcal{S}\, \mathcal{A}_{\mathcal{U}}\,|0^{\otimes n}\rangle + \langle 0^{\otimes n}|\,\mathcal{S}\,\mathcal{A}_{\mathcal{U}}^{\dagger}\, \mathcal{S}\, \mathcal{A}_{\mathcal{U}}^{\dagger}\, \mathcal{S}\, \mathcal{A}_{\mathcal{U}}\, \mathcal{S}\, \mathcal{A}_{\mathcal{U}}\,|0^{\otimes n}\rangle \Big)^{2} \Big] \, .
\end{align}
Eq.~\eqref{eq:QFI m=2 supp} then depends on the OTOCs calculated in Sec.~\ref{sec:Calculation Details}, and on higher order OTOCs. Indeed, this suggests that, to obtain information about higher order OTOCs through the susceptibility of the system, one has to apply a concatenation procedure: (i) first one has to apply the protocol for $m=1$ to obtain information about the OTOC presented in Sec.~\ref{sec:A Brief Overview of Out-of-Time Order Correlations}, and (ii) one has to apply the protocol for $m=2$ to obtain information about higher order OTOCs. The procedure may eventually continue up to the desired order of OTOCs to be estimated. The present procedure is a suitable alternative to the standard procedure for calculating OTOCs \cite{google2025observation}, as it relates information scrambling to the system's susceptibility, thereby providing a suitable experimental benchmark for the indirect evaluation of OTOCs.

\section{Additional Remarks}
\label{sec:Additional Remarks}

To conclude this Supplementary material, we provide additional calculations and clarifications about the protocols used in the main text. In particular, we explicitly calculate the ``randomness" of the state $\ket{\chi} = \mathcal{A}_{\mathcal{U}} \ket{0^{\otimes n}}$, and the physical motivation of an extra final unitary $\mathcal{U}$ in the perturbative protocol.

\paragraph{A Clarification about the protocol} Throughout the main text and Sec.~\ref{sec:Calculation Details}, we had to explicitly obtain the results for $\ket{\psi(\phi)}$ rather than directly relying on some ensemble averages related to Haar states, because the state $\ket{\chi} = \mathcal{A}_{\mathcal{U}} \ket{0^{\otimes n}}$ is not a Haar random state. For the traceless unitary $\mathcal{A}$ used in the local and global protocols, we can demonstrate this by computing the mean squared overlap between $\ket{\chi}$ and $\ket{0^{\otimes n}}$ 
\begin{equation}
    \label{eq:ensemble average chi supp}
    \underset{{\mathcal{U} \sim H}}{\mathbb{E}} \left[ \abs{ \bra{0^{\otimes n}} \ket{\chi} }^2 \right] = \underset{{\mathcal{U} \sim H}}{\mathbb{E}} \left[ \abs{ \bra{0^{\otimes n}} \mathcal{U}^{\dagger} \mathcal{A} \mathcal{U}\ket{0^{\otimes n}} }^2 \right] = \frac{1}{2^n + 1} \, .
\end{equation}
If the state $\ket{\chi}$ had been a Haar random state, the results would have been
\begin{equation}
    \label{eq:ensemble average psi Haar supp}
    \underset{{ \psi_{H} \sim H}}{\mathbb{E}} \left[ \abs{ \bra{0^{\otimes n}} \ket{\psi_{H}} }^2 \right] = \frac{1}{2^n} \, ,
\end{equation}
where the results are derived from the standard ensemble average of Haar random states \cite{mele2024introduction}. Thus, this overlap moment approaches the Haar-state value as the Hilbert-space dimension increases; this agreement alone does not establish convergence of the full state ensemble to the Haar distribution.

\paragraph{Additional Unitary in the Perturbative Protocol} The results presented are based on two similar protocols: the first and exact one which relies on the physical implementation of a state $\ket{\psi(\phi)} = \left[  \mathcal{M}(\phi) \mathcal{U}^{\dagger} \mathcal{A} \mathcal{U} \right] \ket{0^{\otimes n}}$ and a perturbative protocol based on the creation of the state $\ket{\psi(\phi)} = \left[ \mathcal{U} \mathcal{M}(\phi) \mathcal{U}^{\dagger} \left( \frac{\mathbb{1} + i\mathscr{V}}{\sqrt{2}} \right) \mathcal{U} \right] \ket{0^{\otimes n}}$, where here $\mathcal{S} = (\nicefrac{1}{2})\sum_{j=1}^{n} \mathbb{1}^{\otimes j-1} \otimes \sigma^{z}_{j} \otimes \mathbb{1}^{\otimes n-j}$. As can be seen, the perturbative protocol differs from the exact one mainly in that an additional random unitary $\mathcal{U}$ is applied after a parameter-dependent unitary $\mathcal{M}(\phi)$. Calculating the QFI in the perturbative protocol, we obtain $\mathcal{I}(\phi) = 4 \left( \langle \mathscr{A}_{\mathcal{U}}^{\dagger} \mathcal{S}^2 \mathscr{A}_{\mathcal{U}} \rangle - \langle \mathscr{A}_{\mathcal{U}}^{\dagger} \mathcal{S} \mathscr{A}_{\mathcal{U}} \rangle^2 \right) = \frac{4}{s^2} \left( \langle (\mathcal{S}^{\dagger})^2 \mathscr{A}_{\mathcal{U}}^{\dagger} \mathcal{S}^2 \mathscr{A}_{\mathcal{U}} \rangle - \langle \mathcal{S}^{\dagger} \mathscr{A}_{\mathcal{U}}^{\dagger} \mathcal{S} \mathscr{A}_{\mathcal{U}} \rangle^2 \right)$, where here $\mathscr{A}_{\mathcal{U}} = \mathcal{U}^{\dagger} \left( \frac{\mathbb{1} + i\mathscr{V}}{\sqrt{2}} \right) \mathcal{U}$. This equation has the same structure as Eq.~\eqref{eq:QFI pure state protocol supp}, and the additional unitary, being parameter-independent, does not change the form of the QFI; it is only related to the optimal final measurement used to characterize the system \cite{kobrin2024universal}.